\RequirePackage{fix-cm}
\documentclass[twocolumn,epjc3]{svjour3}  
\smartqed  
\journalname{Eur. Phys. J. C}

\usepackage[dvips,dvipdf]{graphicx}

\usepackage{xcolor,amssymb}
\usepackage{color}
\usepackage{epsf}
\usepackage{soul}
\usepackage{etex}
\usepackage{morefloats}

\colorlet{darkgreen}{green!50!black}
\colorlet{brightyellow}{yellow!75!red}
\colorlet{orange}{red!50!yellow}
\colorlet{darkblue}{blue!60!black}
\colorlet{darkred}{red!80!black}

\begin{document}

\title{Protonium annihilation densities in a unitary coupled channel model}

\author{Emanuel Ydrefors\thanksref{e1,addr1}
\and
Jaume Carbonell\thanksref{e1,addr2}  
}
\thankstext{e1}{e-mail: ydrefors@kth.se}
\thankstext{e2}{e-mail: carbonell@ipno.in2p3.fr}
\institute{Instituto Tecnol\'ogico da Aeron\'aautica, DCTA, 12.228-900 Sao Jos\'e dos Campos, Brazil\label{addr1}
\and
Universit\'e Paris-Saclay, CNRS/IN2P3, IJCLab, 91405 Orsay, France\label{addr2}
}

\date{Received: date / Accepted: date}

\maketitle

\today

\bigskip
\begin{abstract}
We consider a unitary coupled channel model to describe the low energy proton-antiproton 
scattering and the lower Coulomb-like protonium states. 
The existence of deeper quasi-bound states of nuclear nature is found to be a consequence of the experimental data. 
The properties of these states as well as the protonium  annihilation densities are described and the difference with respect to the optical models description are manifested.
\keywords{Low energy antiproton physics \and Optical models \and Unitary coupled channel models \and Quasi-nuclear states \and Protonium \and Annihilation}
\end{abstract}

\section{Introduction}

Several experiments in low energy antiproton physics are based on  trapping low energy antiprotons on Coulomb orbits around a nucleus.
These processes take usually place at highly excited Rydberg states and, after being captured, the antiprotons cascade down to the lowest Coulomb  orbits until they are annihilated with the nucleons.
They were experimentally studied in LEAR   \cite{Amsler_ANP18_87,Gastaldi_NPA478_1988}  until its closure in 1996
 and they are also scheduled  in the  foreseen PUMA project \cite{PUMA_LI_2017,PUMA_CERN_Aproval_2021}, both at CERN.

The strong \={N}N force results into shiftenning and broadening the  antiproton-Nucleus (\={p}A) Coulomb states. Their energy
becomes complex $E=E_R- i\;{\Gamma\over2}$ and the differences with respect to the pure Coulomb ones $E_C$  
\begin{equation}\label{DE}
 \Delta E \equiv E-E_C \equiv \epsilon_R - i\;{\Gamma\over2} 
\end{equation}
 have been measured in several experiments and for a large variety of nuclei.
This quantity is a privileged open door for studying and parameterizing the \={N}N interaction. To this aim
a sizeable amount of low energy scattering data -- elastic, annihilation and charge-exchange (\={p}p$\to$\={n}n) cross sections -- 
as well as the level shifts ($\epsilon_R$) and widths ($\Gamma$) of the  lowest   \={p}p (protonium)  levels, essentially S and P states,  were collected during the LEAR period.
The interested reader can find  a summary of the most remarkable achievements  in the last reviews \cite{A_NPA658_99,REV_LEAR_KBMR_PREP368_2002,REV_LEAR_KBR_PREP413_2005,REV_NNB_Frontiers_2020}
and references therein.

Of particular interest is the width of a state $\alpha$,  $\Gamma_{\alpha}$, where $\alpha=\{L,S,J\}$ denotes the ensemble of its  quantum numbers. 
In the existing theoretical models it can be expressed as an integrated quantity over some spatial distribution in the form
\begin{equation}\label{PAD}
 \Gamma_{\alpha}= \int_0^{\infty}  \gamma_{\alpha}(r) \; dr 
 \end{equation}
where $r$ is the distance between the orbiting antiproton and the center of mass of the nucleus. The integrand $\gamma_{\alpha}(r)$, denoted
annihilation density, can be interpreted as the probability for an antiproton, captured  on a given state $\alpha$, to be annihilated at  distance $r$. 
The knowledge of the annihilation density is  relevant in the PUMA project, whose  main interest is
to correlate the annihilation process with the peripheral density distribution of neutrons in the target nucleus.

Most of the existing N\={N} interaction  models are adjusted to reproduce the measured
low energy scattering observables \cite{REV_NNB_Frontiers_2020} and provide reasonable agreement with  the experimentally known protonium $\Delta E$'s  \cite{Ausberger99}.
However the short range part of the p\={p} wave function, determining  $\gamma_a(r)$,  could be strongly dependent on the
physical input of the model, in particular the way it mimics the annihilation dynamics.

The aim of the present work is to compare  the predictions for  $\gamma_a(r)$ in the simplest case  of protonium,  in fact  a coupled  $\bar pp-\bar n n$  system,
as  they  are obtained by the two main families of theoretical approaches to NN interaction: optical models  (OM) and unitary coupled channel models (UCCM). 
The protonium annihilation densities have been computed in the past  using some selected OM \cite{CIR_ZPA334_1989} but no any 
realistic result is known in the framework of  UCCM. 

We will describe in Section \ref{Th_Models} the main contents of the theoretical \={N}N models.
Section \ref{Results} will be devoted to describe some results with the unitary model concerning
\={N}N quasi-nuclear states and S-wave protonium level shifts as well as the corresponding annihilation densities.
Section \label{Conclusion} contain some concluding remarks.

\section{Theoretical models}\label{Th_Models}
 
For the \={N}N interactions that we have  considered in this work, the real part of the \={N}N potential (denoted $U_{\bar NN}$) is
given by the G-parity transform of an NN model  (denoted $V_{NN}$) regularised below some cut-off radius $r_c$.
The $V_{NN}$'s on which we are interested  are in their turn based in meson  exchange theory, that is they result from
the coherent sum of contributions coming from one-meson exchange potentials $V_{NN}^{(\mu)} $ where $\mu=\pi,\rho,,\omega,..$ 
\begin{equation}\label{UNN}
 V_{NN}= \sum_{\mu}  V_{NN}^{(\mu)}   
\end{equation} 

For $r>r_c$,   $U_{\bar NN}$ takes the form 
\begin{equation}\label{UNNb}
 U_{\bar NN}(r>r_c)  = \sum_{\mu=\pi,\rho,\omega,...} \;  G(\mu)\; V_{NN}^{(\mu)}(r) 
 \end{equation} 
where G($\mu$) is the G-parity of meson $\mu$. G is related to the charge-conjugation (C) and isospin (T) quantum numbers by  $G= C (-)^T$.

For $r<r_c$, a regularization procedure is needed to avoid  the non integrable short range singularities (essentially attractive $1/r^2$ and $1/r^3$ terms)
due to the spin orbit and tensor terms. This procedure depends on the particular model. 

Thus, the real part of the different \={N}N models differ from each other in the choice of their meson content 
\cite{DR1_PRC21_1980,DR2_RS_PLB110_1982,KW_NPA454_1985,CLLMV_PRL_82,PLLV_PRC50_1994,ELLV_PRC59_1999,PARIS_PRC79_2009,NNB_CC_Nijm_1984} 
as well as in the regularization procedures.
It is worth noticing however, that during the last years, successful EFT inspired N\={N} potentials  have been built \cite{Haidenbauer_JHEP_2014,Haidenbauer_JHEP_2017}  keeping only the pion
as the explicit exchanged meson and replacing the heavier ones by contact terms.
They are naturally formulated in momentum space and have a strong non local character which make it difficult to be used
in configuration space where the annihilation densities are defined. We will not consider them at this stage of our research.

The main differences among the models rely however  on the way they account for the annihilation process,
giving rise to two main families: optical models (OM) and unitary coupled (UCCM) channel models that we will briefly describe in what follows.

\subsection{Optical models}\label{O_Models}

In this approach, the annihilation is  described by adding to $U_{\bar N N}(r)$ a complex potential $W(r)$ such that
\begin{equation}\label{VOM}
 V_{N\bar N}(r)=  U_{\bar N N}(r)+  W(r) 
 \end{equation}

The imaginary part of $W$ accounts for the loss of flux in the p\={p} channel due to annihilation.
In the simplest cases -- Dover-Richard  version 1  \cite{DR1_PRC21_1980} (DR1), version 2  \cite{DR2_RS_PLB110_1982} (DR2) and Kohno-Weise   \cite{KW_NPA454_1985} (KW)  -- 
$W$ is local, energy- and state-independent, and have the common form
\begin{equation}\label{W}
 W(r)=- {W_0\over 1+ e^{ r-R\over a} }  
\end{equation} 
with the parameters given in Table \ref{Tab_Par_OM}.
\begin{table}[h]
\begin{center}
\begin{tabular}{l l l  l}
            &     DR1       &   DR2    &  KW  \\\hline
W$_0$ (GeV) &    21+20i   &  0.5+ 0.5i             &  1.2i       \cr
R  (fm)        &      0          &    0.8           &   0.55     \cr
a   (fm)         &     0.2        &    0.2           &   0.2       
\end{tabular}
\caption{Parameters (GeV and fm) of the Dover-Richard (DR1 and DR2 versions)  and Khono-Weise (KW)  \={N}N optical models.}\label{Tab_Par_OM}
\end{center}
\end{table}
Although differing either by their meson contents or by their annihilation potential $W$, these three models
 give similar results for the protonium S- P- and D- waves \cite{CIR_ZPA334_1989}.

\begin{figure}[!ht]
\begin{center}
\includegraphics[width=7.cm]{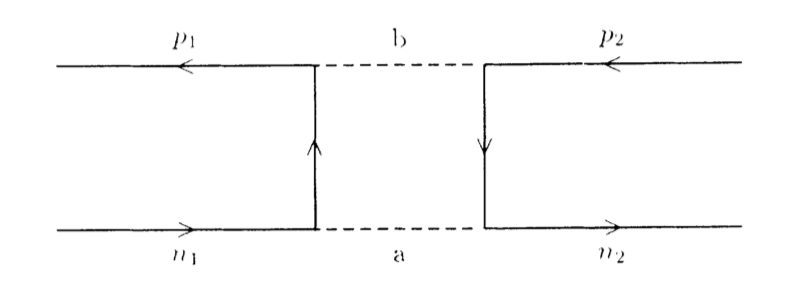}
\caption{\={N}N annihilation into two mesons ($ab$) and re-annihilation into \={N}N}\label{W_Paris}
\end{center}
\end{figure}

The Paris N\={N} potential  \cite{CLLMV_PRL_82,PLLV_PRC50_1994,ELLV_PRC59_1999,PARIS_PRC79_2009}  provides a more elaborate non-local, state- and energy-dependent  version of $W$ 
 based on the two-meson annihilation and re-annihilation process despicted in Figure \ref{W_Paris} that was explictly computed in \cite{Bachir_PhD_Paris_1980}.
 Taking into account this diagram results into a complex annihilation potential $W$ acting in the N\={N} channel with a characteristic range of
\[  r_W={1\over 2M_N}  \approx 0.1 \;{\rm fm}\]

The success of OM to describe the phenomenology of  low energy p\={p} physics is remarkable \cite{REV_NNB_Frontiers_2020}.
However the use of complex potentials implies a non hermitian Hamiltonian and a  non unitary S-matrix with $\mid U^{\dagger}U \mid <1$.
This produces the loss of some standard symmetry properties of the S-matrix and introduce  some unphysical properties in the description of bound states and resonances.
These drawbacks, first pointed out in \cite{KKMS_JETPL26_1977,SH_PR35_1978},  have been examined in  \cite{CDPS_NPA535_91} in the case of KW N\={N} optical potential.

The strong part of NN and N\={N} force is derived in the isospin  basis. 
However this basis is not adapted for computing low-energy p\={p} scattering processes and/or the protonium bound states due to the relevant
role of Coulomb interaction and, to a less extend, the p-n mass difference $\Delta_{np}=2(m_n-m_p)$  \cite{CIR_ZPA334_1989}.
We used instead the so called particle-basis where the  $p\bar{p}$ and $n\bar{n}$ states are coupled  by the "charge-exchange potential".
They are described by a  two-channel Schrodinger equation 
\begin{equation}\label{SCHEQ_OM}
 (E- H_0 ) \Psi= \hat{V}\;  \hat\Psi   
 \end{equation}
with
\[ \quad \hat\Psi=  \pmatrix{   \Psi_{ p\bar{p}}   \cr \Psi_{n\bar{n}}  }     \quad 
\hat{V}= \pmatrix{
V_{p\bar{p}}     & V_{ce}      \cr
V_{ce}             &       V_{n\bar{n}}                 }
\]
and a channel-diagonal kinetic energy term
\begin{equation} 
H_0 = - {\hbar^2\over M} \Delta   \qquad  M={m_p+m_n\over2} 
\end{equation}
Using the isospin conventions \cite{Gasiorowicz_1966,CIR_ZPA334_1989} 
\begin{equation}\label{Nbar_doublet}
N = {p\choose n}  \quad   \bar{N} = {- \bar n\choose + \bar p}   \equiv  
\begin{array}{lcl}
|1/2,+1/2> &=& -|\bar n> \cr
|1/2,-1/2>  &=& +|\bar p>
\end{array}
\end{equation}
the particle basis is expressed in terms  of N\={N}  isospin  states  $\mid T,T_3>$  as
\begin{equation}\label{particle_basis}
\begin{array}{lcl c lcl}
|p\bar{p}>&=&+{1\over\sqrt2}\left\{|00> + |10>\right\}  \\
|n\bar{n}>&=&+{1\over\sqrt2}\left\{|00> - |10>\right\}  \\
|p\bar{n}>&=&-|1,+1>                                 \\
|\bar{p}n>&=&+|1,-1>           
\end{array}
\end{equation}
and the  matrix elements of  $\hat{V}$  read
\begin{eqnarray}
V_{p\bar{p}}  &=& {1\over2} \left( {V^{T=0}_{N\bar{N}} + V^{T=1}_{N\bar{N}} }  \right)   + V_C   \cr
V_{n\bar{n}}  &=&  {1\over2} \left( {V^{T=0}_{N\bar{N}} + V^{T=1}_{N\bar{N}} }   \right)+  \Delta_{np} \cr
 V_{ce}          &=&  {1\over2} \left( {V^{T=0}_{N\bar{N}} - V^{T=1}_{N\bar{N}} }  \right)     \label{Vpb_OM}
\end{eqnarray}
where $V_{n\bar{n}}$ incorporates the p-n mass difference and $V_{p\bar{p}}$  the Coulomb $p\bar{p}$ interaction 
\begin{equation}\label{VC}
V_C= - { \alpha(\hbar c) \over r} 
\end{equation}

For a protonium state $\hat\Psi$ with complex energy E, the annihilation density  (\ref{PAD}) is given by
\[ \gamma=\sum_{c=p\bar{p}, n\bar{n}} \gamma_{c}  \]
where\footnote{Notice that a factor 2 is missing in eq (1) of  \cite{CIR_ZPA334_1989} } 
\begin{equation}\label{PAD_OM}
 \gamma_{c} = - 2 \; {\rm Im} [W_c(r)]   \mid u_c(r) \mid^2  
 \end{equation}
is the $p\bar{p}$ and $n\bar{n}$  contributions and $u_c(r)$ is the corresponding normalised reduced radial wave function, solution of (\ref{SCHEQ_OM}).
This expression  follows  from the very definition of a complex  eigenstate
\begin{equation}\label{IE_OM}
E = E_R-i{\Gamma\over 2} =  < \Psi \mid H_0 + V \mid \Psi > 
\end{equation}
and the fact that the non hermitian part of the hamiltonian is ${\rm Im}(W)$. This gives  
\begin{equation}\label{IGamma_OM}
 {\Gamma\over 2}= -   \left\{ < \Psi \mid   {\rm Im} (W) \mid \Psi >  \right\}  
 \end{equation}

\subsection{Unitary coupled-channels model (UCCM)}\label{UCC_Models}

\begin{figure}[h!]
\begin{center}
\includegraphics[width=5.cm]{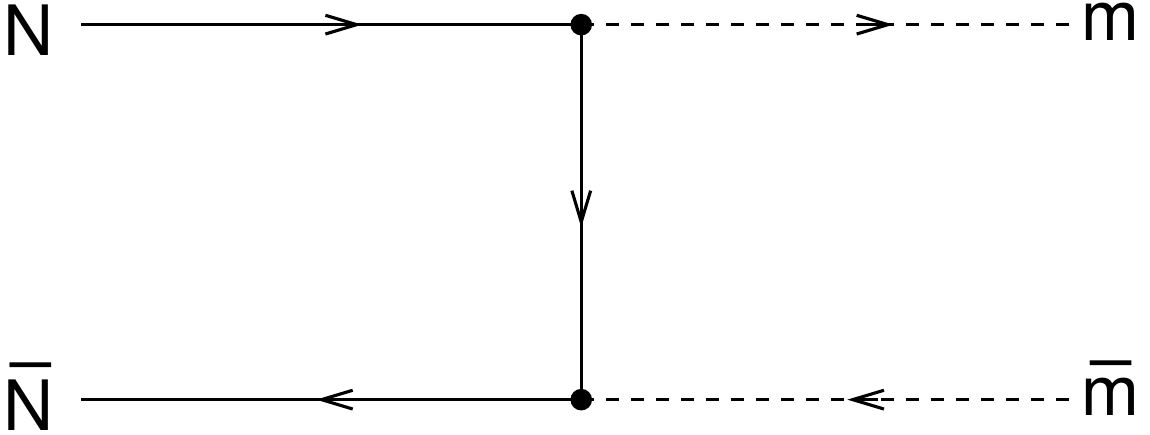}
\caption{\={N}N annihilation into two mesons ($m \bar m$)}\label{NNB_annihil}
\end{center}
\end{figure}

This family of models  aims to preserve the hermiticity
of the full hamiltonian -- and consequently the unitarity of the S-matrix -- by  coupling each N\={N} isospin component  to an explicit meson-meson ($m\bar m$) channel
\begin{equation}\label{NN_T}
\mid N\bar{N}\rangle_T \qquad \to \qquad   \mid N\bar{N}\rangle_T= \pmatrix{   \Psi^T_{ N\bar{N}}   \cr \Psi^T_{m\bar{m}}  } 
\end{equation}
This   N\={N}$\to$ m\={m}  coupling  is realised with a short range  Yukawa like potential 
\begin{equation}\label{Va}
 V_a^{T}(r) = \lambda_T\;(\hbar c)\; {e^{-{r\over r_a}}\over r}   
 \end{equation}
 corresponding to the diagram of Fig. \ref{NNB_annihil}  and which is,  abusively, denoted "annihilation potential".

The range ${r_a}$  of  the annihilation potential   (\ref{Va})  thus defined, is given  by the nucleon Compton wavelength 
\begin{equation}
r_a={1\over m_p}\approx{0.21\ fm}
\end{equation}
This  value is in agreement with the views of Refs. \cite{MARTIN 61,SH_PR35_1978}  which
define the annihilation range as the nearest singularity  of the scattering amplitude (represented in Fig 1) in the $t$-channel.
They obtain a lower limit $r_a<1/2m_p$, which is independent of the number of annihilating mesons  along the $t$-channel.
It is also in agreement  with the characteristic range $r_W$ of  the Paris annihilation potential \cite{Bachir_PhD_Paris_1980}.
The question of "annihilation range" has been abundantly debated in the past;
the interested reader can  profitably consult  \cite{Erice_1988,Mainz_1988,Haidenbauer_ZPA334_1989,REV_NNB_Frontiers_2020} and references therein.

This provides for each isospin state, a  two-channel Schrodinger equation
\begin{equation}\label{SCHEQ_T}
(E-H_0)  \pmatrix{\Psi^T_{ N\bar{N}} \cr \Psi^T_{m\bar{m}}} 
=\hat{V}  \pmatrix{ \Psi^T_{N\bar{N}} \cr \Psi^T_{m\bar{m}}  }    
\end{equation}
where 
\begin{equation}\label{VCC}
 \hat{V}^T =\pmatrix{ V_{N\bar N}^T  & V^T_a \cr  V^T_a & V^T_{m\bar m}  - \Delta  }
 \end{equation}
 is a real and hermitian potential matrix and $\Delta=2(M-M_{m})$ accounts for  the threshold mass difference
 between the nucleon-nucleon  (NN) and meson-meson (m\={m}) channels.
 We will assume hereafter that    $V_{m\bar m}=0$, so that mesons  interact only via the coupling to N\={N} channel. 

The main difference with respect to optical models is that, even for bound states (with Re(E)$>-\Delta$), the second channel is kinematically open  and has scattering boundary conditions.
This is better illustrated by writing equation (\ref{SCHEQ_T}) in terms of the reduced radial solutions $u_c(r)$ 
\begin{eqnarray} 
{\hbar^2\over2\mu_1}u_1''+ \left[E\; -\; {l_1(l_1+1)\hbar^2\over2\mu_1r^2} - V_{11} \right]u_1&=&V_{12}u_2\cr
{\hbar^2\over2\mu_2}u_2''+ \left[E\; -\; {l_2(l_2+1)\hbar^2\over2\mu_2r^2} + \;\Delta\;  \right]  u_2&=&V_{21}u_1   
\end{eqnarray} 
where $\mu_c$ and $l_c$ denotes the reduced mass and angular momentum of channel $c=1,2$.

For the S-wave, the asymptotic boundary conditions of channel $u_c$ are given by
\begin{equation}\label{ASYMP}
 u_c(r)=e^{iq_c r} 
 \end{equation}
where $q_c$ are the channel momenta, related to the energy of the state by
\begin{eqnarray}
q_1^2 &=& {2\mu_1\over \hbar^2} E \cr
q_2^2 &=& {2\mu_1\over \hbar^2} (E + \Delta) 
\end{eqnarray}

\begin{figure}[h!]
\begin{center}
\includegraphics[width=8.cm]{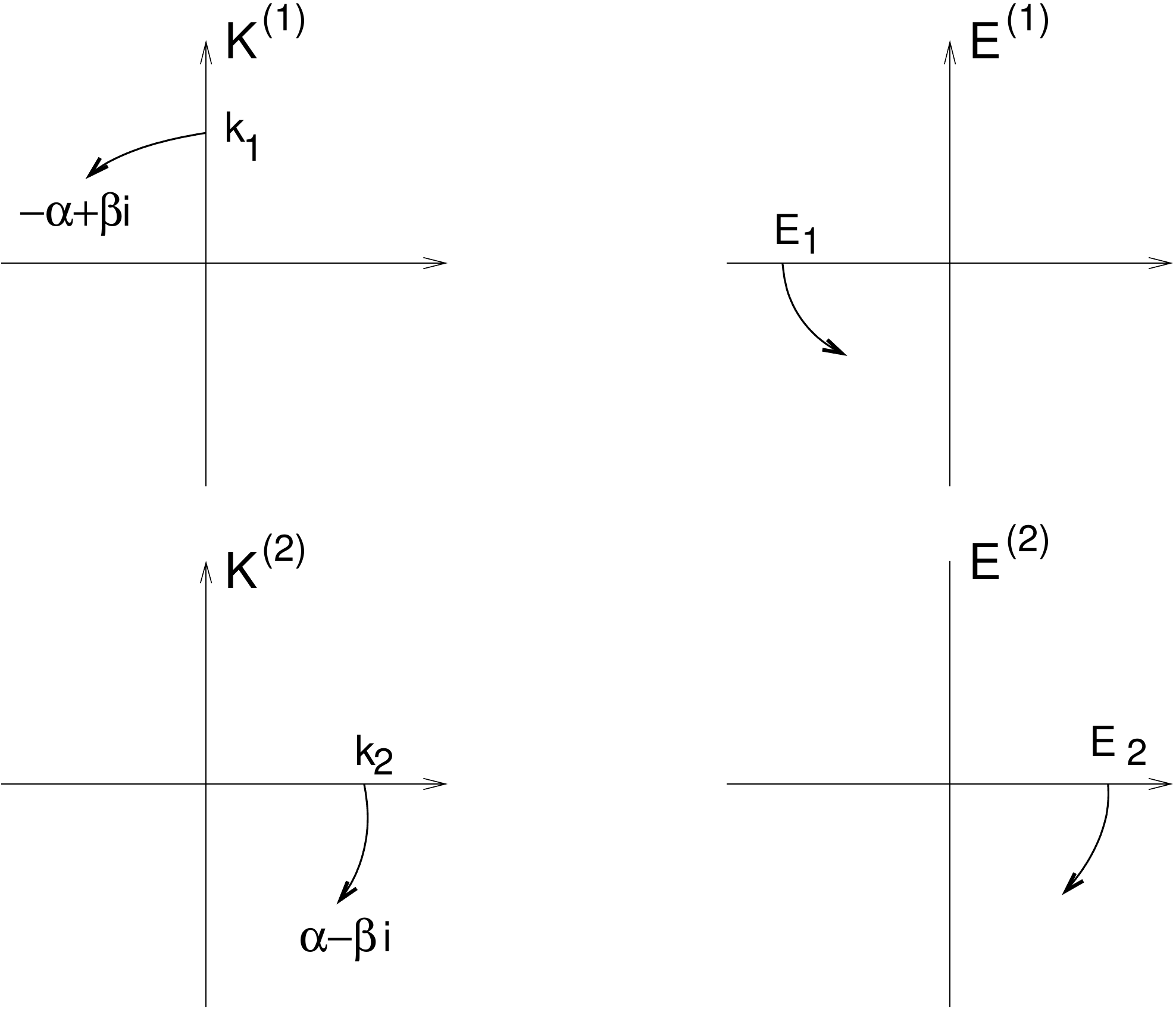}
\caption{Complex momentum ($K^i$)and Energy ($E^i$) sheets  corresponding to protonium states in the two channels of the unitary coupled-channel model (\ref{SCHEQ_T}) }\label{K_E_Sheets_CC2}
\end{center}
\end{figure}

The different choices for the complex squared root define a multivalued complex momentum (K)  and energy (E) manifolds.
For the two-channel problem they are represented schematically in Fig. \ref{K_E_Sheets_CC2},
 with,  the following determinations of the complex squared root :
\begin{itemize}
\item For the first channel with  $-\Delta<{\rm Re}(E) <0$ and ${\rm Im}(E)\le 0$, $q_1$ is defined with the  squared root determination  such that ${\rm Re}(q_1)\le 0, {\rm Im}(q_1)>0$,
 i.e. 
\begin{equation}\label{q1}
   q_1= -\alpha_1+ i \beta_1 \quad e^{iq_1r}=e^{-i\alpha_1r} e^{-\beta_1r}    
\end{equation}   
It corresponds to a squared integrable solution.

\item For the second channel with ${\rm Re}(E+\Delta)>0$, $q_2$ is defined with  ${\rm Re}(q_2)>0, {\rm Im}(q_2)<0$\par
It corresponds to a scattering resonant state, that is exponentially increasing\par
\begin{equation}\label{q2}
 q_2= \alpha_2-  i \beta_2     \quad e^{iq_2r}=e^{i\alpha_2r} e^{\beta_2r}     \quad \alpha_2,\beta_2>0 
\end{equation} 
\end{itemize}
where $ \alpha_1,\beta_1>0$.
A more detailed explanation of the non trivial analytical structure of the  K- and E-manifolds corresponding to a coupled channel model can be found in \cite{BMS_SJNP30_1979}. 

\bigskip
The annihilation densities can be obtained by means of a similar expression than for optical models.
The formal derivation is however more involved and has been detailed in   \ref{App_PAD}.

The final result  for the  width,    is expressed in terms of the annihilation channel wavefunction only
\begin{equation}\label{GammaT_CC}
{ \Gamma\over 2} =  \left\{    {\hbar^2\over2\mu_2}  \int_0^R dr\; {\rm Im} [ u^*_2(r) u''_2(r) ] \right\} 
 \end{equation}
and the corresponding annihilation density is given by
\begin{equation}\label{LAD_CC}
\gamma_a(r)=   2 \; \left\{    {\hbar^2\over2\mu_2} \; {\rm Im} [ u_2^*(r) u''_2(r) ] \right\}  
 \end{equation}
which constitutes the UCCM counterpart of  (\ref{PAD_OM}). 

The relevance of an hermitian approach to the  \={N}N problem 
was put forward  before starting  the LEAR era, by  I.S. Shapiro and collaborators, from  the ITEP and Lebedev groups  in Moscow  
\cite{KKMS_JETPL26_1977,SH_PR35_1978,BMS_SJNP30_1979}.  
The main reasons  were {\it (i)} to keep  the right analytical properties of the S-matrix which are lost  in the optical models due to the
non hermitian character   and {\it (ii)}  to have a non vanishing short range $\bar p p$ wave function which,  due to the strong imaginary part of the optical models, is extremely suppressed.
Both facts  --  together with a short range annihilation potential --  have dramatic consequences in the properties  (energies and widths) of the S-matrix  poles associated
with the  bound states \cite{CDPS_NPA535_91}  and in the description of the annihilation process itself.
These differences affect both the deeply quasi-bound and resonant states of nuclear origin (the so called "quasi-nuclear states", first proposed in \cite{DMS_JETPL10_1969,DMS_SJNP11_1970,DMS_NPB21_1970,DMS_PMNP_1971}), as well as the loosely bound Coulomb-like states of protonium.

\begin{figure}[h!]
\begin{center}
\includegraphics[width=9.cm]{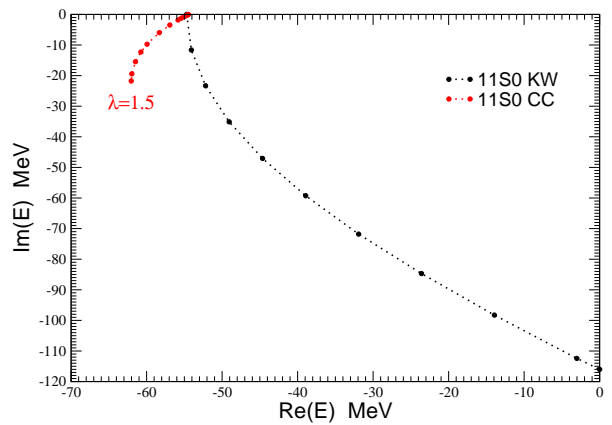}
\end{center}
\vspace{-.5cm}
\caption{Complex energy trajectory of a $^{11}$S$_0$ state as a function of the annihilation strength in optical a unitary coupled channel models}\label{Pole_Traj_11S0_QN}
\end{figure}

As an  illustrative example  of such a different behaviour we have compared in Figure  \ref{Pole_Traj_11S0_QN} the  complex energy  trajectories of a $^{11}$S$_0$  bound state as a function of the annihilation strength
obtained in  an optical model and in a unitary coupled channel model. 
This is a pure isospin (T=0) state, generated by the strong \={N}N interaction only,  hence its "quasi-nuclear" denotation.
The results for the optical model (black dots) are taken from  \cite{CDPS_NPA535_91} and correspond to KW model. 
In absence of annihilation ($W_0$=0) it is a bound state with energy E=-54.7 MeV and rms radius  R=$\sqrt{<r^2>}$=1.3 fm. 
This energy, that become complex for $W_0\ne 0$,  is plotted as a function of $W_0$ defined  in (\ref{W}). 
We have adjusted the parameters of the UCCM  described in the next section, in order to locate a bound state at the same energy and vary the annihilation strength $\lambda$ in the physical domain.
The corresponding trajectory $E(\lambda)$ is indicated in red dots.

In OM, the  annihilation has always a "repulsive" effect, that is   decreases the real part of the bound state  energy and "pull out" the state towards the continuum; the threshold is crossed at $W_0\approx0.5$ GeV, 
well before its model value $W_0=$1.2 GeV.
In a unitary model the situation is totally opposite: in the range of parameters of interest the annihilation has an attractive result, with
an increasing  energy  Re(E). Notice also that the level shift and the width of the state remains much smaller than for OM. 

As it was  shown in refs  \cite{KKMS_JETPL26_1977,SH_PR35_1978},  
the origin of this striking differences  --  small level shifts and width and effective attraction in the \={N}N channel  induced by the annihilation --  
relies ultimately in two model properties: its unitarity  and  the short range of the annihilation potential.
On one hand,  while in the optical model the \={p}p wave function is systematically suppressed (loss of flux in the \={p}p channel that goes nowhere),
 in the UCCM it is, at the same time and with equal probability, populated by the "re-annihilation process",  i.e. by the reverse  \={n}n$\to$\={p}p coupling. 
For small or moderate value of the annihilation strength, the net effect is attractive and the corresponding \={p}p wave function can be even enhanced
with respect to the free case.  
On the other hand,  if the size of the state is larger than the annihilation range  ($r_a<<R$) the level shifts and widths of the state 
are loosely affected by the coupling and remain small.

This example concerns a relatively deep, quasi-bound state of quasi-nuclear nature.
We will see in the next section other examples taken from the Coulomb-like protonium states, involving energies several order of magnitude smaller
and much larger sizes,  where the complex energy trajectories have a different behaviour.

\vspace{-.5cm}

Exhausting the \={N}N annihilation  process by a fictitious two-meson channel,
furthermore taken interaction free $V_{m\bar m} \equiv0$, is an 
oversimplification of the physical problem, involving even at zero energy dozens of annihilation channels with many particles in the final state.
It is however a consistent way to guarantee the unitarity of the \={N}N model  and the proper analytic structure of the scattering amplitudes.
The unitarity relies on the use of a real transition potential (\ref{Va})  inserted in the symmetric potential matrix  (\ref{VCC})  of  an  hermitian equation (\ref{SCHEQ_T}).

This fact and the existence of two different interaction ranges ($r_N\sim1$ fm in the N\={N} chanel, $r_a\sim0.2$ fm 
in the N\={N}$\rightarrow$m\={m} transition potential), constitute the key points of this approach.
The UCCM approach has been successfully used in many problems 
related to baryon-antibaryon states  \cite{DPS_IJMPA5_1990,DCP_SJNP52_1990,CPD_NP558_1993,CPD_PLB306_1993,DCK_NCA107_1994} 
but it has never been applied  to describe with realistic interactions the level shifts of protonium low lying states, or equivalently the $p\bar{p}$ scattering lengths.
This will be done in the next section for a particular choice of $V_{N\bar N}$.

\section{Results from a unitary coupled-channels model}\label{Results}

\subsection{The model}\label{The_Model}

The  \={N}N unitary  coupled channel model that we have considered  is a modified version (hereafter denoted Model I)  of that used in \cite{DCP_SJNP52_1990,CPD_NP558_1993,CPD_PLB306_1993,DCK_NCA107_1994}
for describing the low energy nucleon-antinucleon interaction at $p_{lab}<300$ MeV/c as well as the $p\bar{p}\to \Lambda\bar{\Lambda}$ nearthreshold phenomena.

The meson contents of the N\={N} interaction --  two pseudoscalar ($\pi,\eta$), two vector ($\rho,\omega$) and two scalars ($\sigma_0,\sigma_1$) -- was taken 
from the pioneering  Brian-Phillips work \cite{BP_NPB5_1968} with an updated $\pi N$ coupling constant. The corresponding masses, quantum numbers and couplings are
given in Table \ref{Table_mesons}. 
\begin{table}[h!]
\[\begin{array}{  c   c   c  c c  c  c  } 
  1      & m     & J^{\pi}  & T & G & g     & f/g  \\   \hline
\pi      & 138.2 & 0^{-}    & 1 & - & 13.5  &  -   \\   
\eta     & 548   & 0^{-}    & 0 & + & 7.0   &  -   \\   
\rho     & 760   & 1^{-}    & 1 & + & 0.68  &  4.4 \\  
\omega   & 782   & 1^{-}    & 0 & - & 21.5  &  0   \\  
\sigma_0 & 770   & 0^{+}    & 1 & - & 6.1   &  -   \\   
\sigma_1 & 560   & 0^{+}    & 0 & + & 9.4   &  -      
\end{array}\]
\caption{Meson coupling constants for the $\bar{N}N$ potential}\label{Table_mesons}
\end{table}

The original model, formulated in  the isospin basis,  has been readjusted  in view of describing the protonium states, with the isospin symmetry  broken by the Coulomb forces and the $pn$ mass difference.
Indeed, it appeared necessary to work from the very beginning on the
p\={p}-n\={n} particle basis and properly account for the n\={n} threshold and Coulomb effects.
These effects, as well as the tensor coupling potential,  were not taken into account in the previous calculations \cite{DCP_SJNP52_1990,CPD_NP558_1993,CPD_PLB306_1993,DCK_NCA107_1994}.
Its importance will be discussed in the next sections.

\bigskip
As it was the case  for optical models the  \={p}p and \={n}n channels are coupled to each other via the charge exchange potential ($V_{ce}$).
Each of them is in addition coupled to a meson-meson channel via the so called annihilation potentials ($V^a$).
The meson-meson interaction is neglected but, as  for the n\={n}  channel (\ref{Vpb_OM}),  the corresponding potential incorporates the threshold mass difference $\Delta$=$2(M_p-M_m)$. 
This results into a real hermitian potential matrix which guarantees the unitarity and has the following structure:
\begin{equation}\label{POT}
\hat{V}=
\pmatrix{V_{p\bar p}  &V_{p\bar p}^a&V_{ce}       &V_{ce}^a     \cr
         V_{p\bar p}^a&   -\Delta        &V_{ce}^a     &0            \cr
         V_{ce}       &V_{ce}^a     &V_{n\bar n}  &V_{p\bar p}^a\cr
         V_{ce}^a     &    0        &V_{n\bar n}^a& -\Delta              }
\end{equation}
where the  components  $V_{p\bar p}$,  $V_{n\bar n}$ and $V_{ce}$  
in (\ref{POT}) are expressed in terms of the isospin components
as for the OM case (\ref{Vpb_OM}) and the annihilation potentials given by
\begin{eqnarray}
V^a_{p\bar p}(r)  &=&  {\lambda_{0}+\lambda_{1}\over2}    \; {e^{-{r\over r_a}}\over r}    \label{Va_ppb}\\
V^a_{ce}(r)         &=&  {\lambda_{0}-\lambda_{1}\over2}   \; {e^{-{r\over r_a}}\over r}    \label{Va_ce}
 \end{eqnarray}

The state vector consist in a {\bf  four channel vector} obeying  a Schrodinger equation
\begin{equation}\label{SCHEQ}
(E-H_0)|\Psi>=\hat{V}|\Psi>  \hspace{1.0cm}   |\Psi>=\pmatrix{ p\bar{p}\cr m\bar{m}\cr n\bar{n}\cr m\bar{m}} 
\end{equation}
in which E is the center of mass energy energy with respect to the p\={p} threshold.

Equation (\ref{SCHEQ}) can be solved in differential form after a partial wave analysis of each channel wave function. This gives for the
reduced radial wave functions $u_{\alpha}(r)$ a system of coupled differential equations
\begin{equation}\label{RSCHEQ} 
u_{\alpha}''+\left[k_{\alpha}^2-{L_\alpha(L_\alpha+1)\over r^2}\right]  u_{\alpha}-\sum_{\beta}v_{\alpha\beta}u_{\beta}=0
\end{equation} 
where
$v_{\alpha\beta}$=${2\mu_{\alpha}\over \hbar^2}V_{\alpha\beta}$ and $\alpha(\beta)$ label  the channel and partial  wave quantum numbers $\{L,S,J\}$,
and the channel momenta $ k_{\alpha}$, defining the asymptotic boundary conditions,  are given by the relation
\[ k_{\alpha}^2 = {2\mu_{\alpha}\over \hbar^2} \; [ E + 2(M_p-M_{\alpha}) ] \]
ensuring  the same  total  c.o.m. energy in all of them $\sqrt{s}=2\sqrt{M_{\alpha}+k_{\alpha}^2}\approx 2M_i - k_i^2/ M_i$.
Notice that when using the differential  form (\ref{RSCHEQ})  the threshold mass differences ($\Delta$'s) must be removed from the potential matrix (\ref{POT}). 

The masses $M_i$ of the different channels are respectively $M_1$=$m_p$=938.28 MeV, $M_2$=$m_n$=$m_p$+1.2933 MeV, $M_2$=$M_4$=763.0 MeV.

We will use all along this paper the standard spectroscopic notation $^{(2S+1)}$L$_J$, eventually completed with isospin $^{(2T+1)(2S+1)}$L$_J$.
For the uncoupled states ($^1$S$_0$, $^1$P$_1$, $^3$P$_0$,...)  the number of coupled Schrodinger equations  (\ref{RSCHEQ}) is 4 
and  the case of tensor coupled states (e.g: $^3SD_1$, $^3PF_2$,...)  this number  is doubled.

\bigskip
The singularities at $r$=0 due to the spin-orbit and tensor terms of  $V_{N\bar{N}}$,  are regularised    by
imposing, below some cutoff radius $r_c$,  a  $C^1$ polynomial matching in the form: 
\begin{equation}\label{VREG}
{V(x)}={V_0} + Ax^{\nu} + Bx^{\nu+1}   
\end{equation}
where $x={\left(r\over{r_c}\right)}$.
The coeficients $A$ and $B$ are determined by the matching conditions.
We have  fixed the value $V_0$=0 to minimise the short range
effects of a potential which is anyway unknown at small distances, and we have examined the  two extreme cases $\nu$=1 and $\nu$=10.
The final result is a smoothed version of the  OBE potential suitable for numerical calculations.
This regularisation procedure is  implemented to  each isospin component of a given partial wave $\alpha\equiv \{S,L,J\}$.

To check the stability of the OBE predictions in   $U_{N \bar N}$, we have also considered an alternative model (denoted Model II)  built 
with the meson contents of the Nijmegen  \={N}N potential \cite{NNB_CC_Nijm_1984} although regularised with the same prescription (\ref{VREG}). 
This model contains a different set of mesons $(\pi,\eta,\eta',\rho,\phi,\omega,\epsilon)$  with slightly different coupling constants. 
In general Model II results into a weaker potentials, specially in the  T=0 channel, due to the stronger cancellations among the mesons.
This is compensated by using a smaller values of $r_c$, but the main conclusions were  found qualitatively the same in both cases.

\vspace{-0.5cm}
\begin{figure}[h!]
\begin{center}
\includegraphics[width=8.cm]{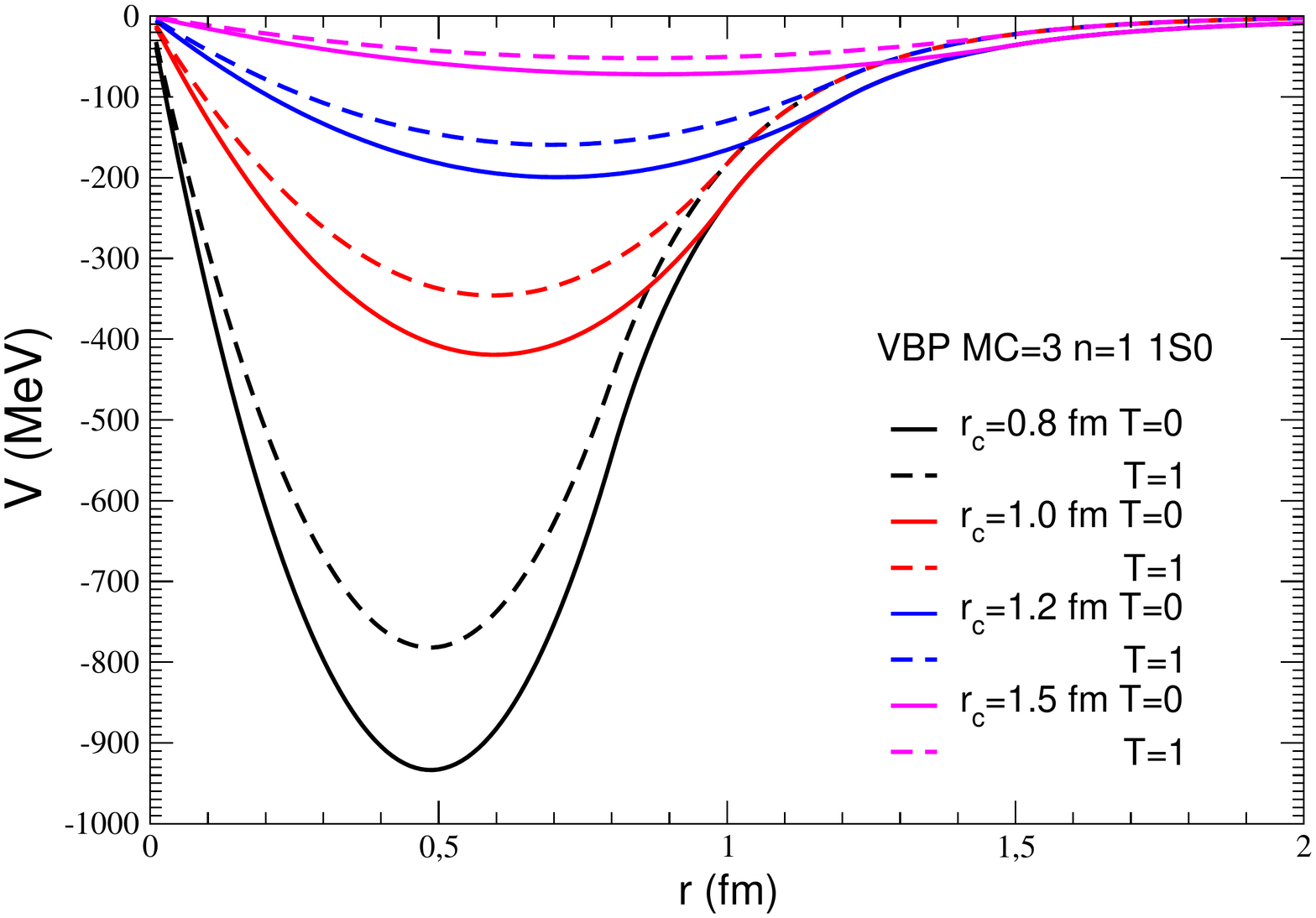} \vspace{-0.cm}
\includegraphics[width=8.cm]{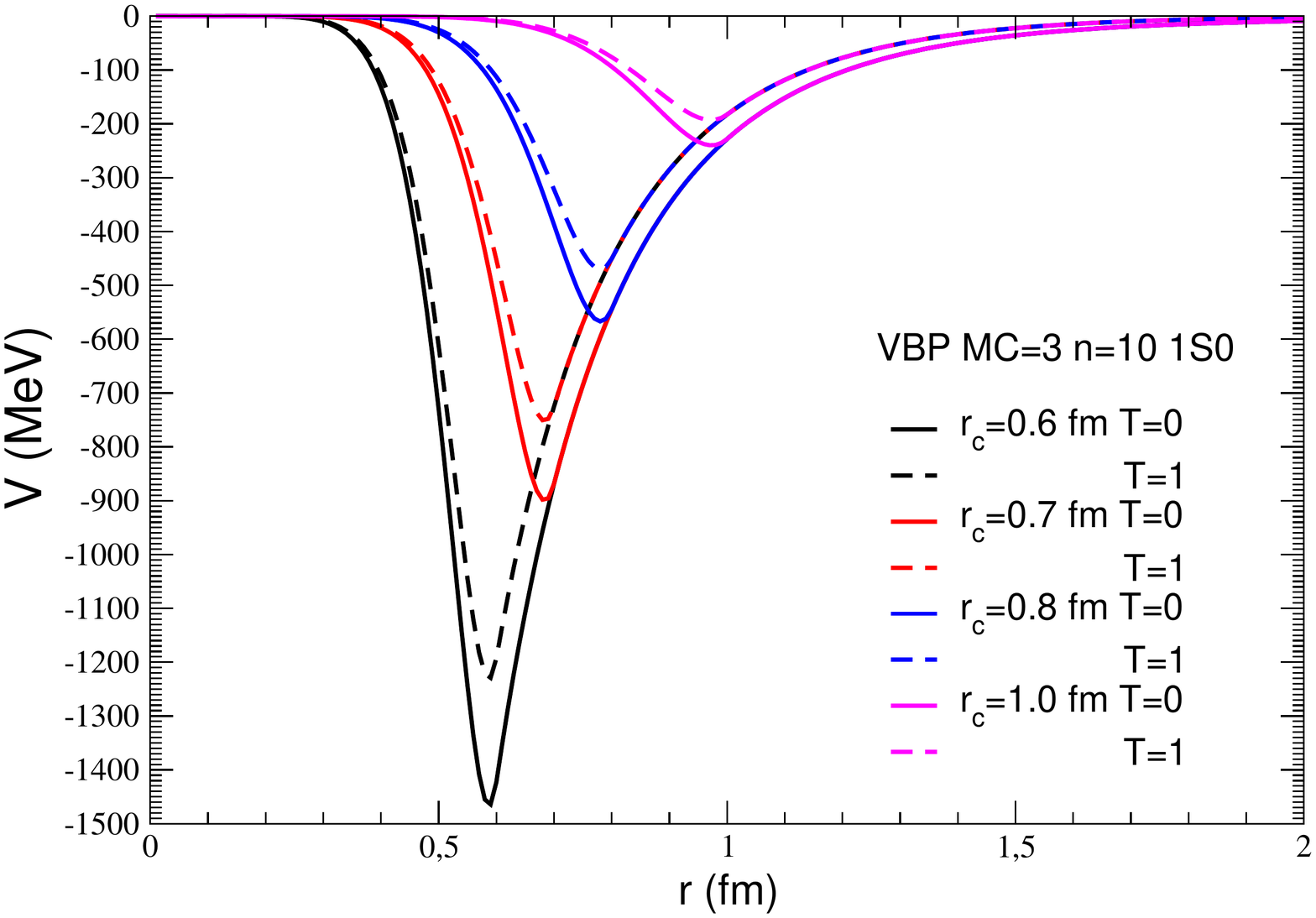}
\end{center}
\vspace{-0.5cm}
\caption{Regularized OBE part of the N\={N} potential in the $^{1}S_0$ partial wave with parameters $\nu$=1 (upper panel) and $\nu$=10 (lower panel ) in (\ref{VREG})
and different values of the  cut-off parameter $r_c$}\label{VNNB_1S0_rc}
\end{figure}

We have plotted in Figure \ref{VNNB_1S0_rc} the isospin components of the N\={N} potential  (Model I) for 
the $^{1}S_0$  partial waves corresponding to  $\nu$=1 (upper panel) and $\nu$=10 (lower panel) in (\ref{VREG}) and for different values of the cut-off parameter $r_c$.
The interaction is reconstructed ''inwards'' through the peripheral meson-nucleon coupling constants, in principle the best well stablished one.
In particular, for $\nu=$10 the interaction becomes negligible below $r\approx r_c$.

\begin{figure}[h!]
\begin{center}
\includegraphics[width=9.cm]{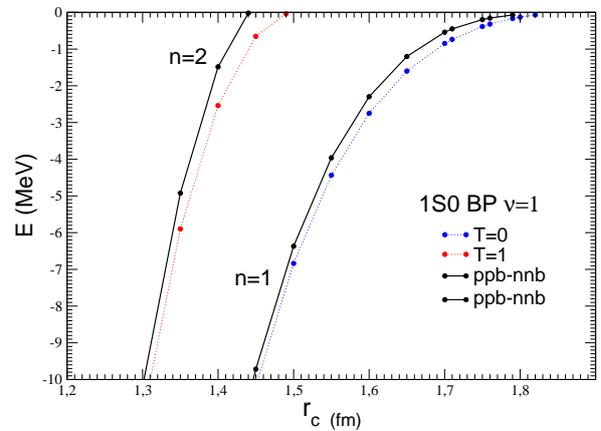}
\end{center}
\vspace{-0.5cm}
\caption{Spectra of the $^1$S$_0$ state with BP potential  (Model I)  with $\nu$=1 as a function of the cut-off radius $r_c$.  
Dashed coloured lines correspond to pure isospin (T) states and black solid lines to the coupled \={p}p-\={n}n ground (n=1) and first excited (n=2) states including Coulomb and $\Delta_{np}$ terms.}\label{B_rc_BP_1S0_n_1}
\end{figure}

The lower quasi-nuclear  spectra of the $^1$S$_0$ state as a function of the cut-off radius $r_c$  is displayed
in Figure \ref{B_rc_BP_1S0_n_1}. It corresponds to Model I (BP) with $\nu$=1. 
Colored lines correspond to the uncoupled isospin states (blue for T=0, and red for T=1)
while the solid black lines to the ground (n=1) and first excited  (n=2) coupled  p\={p}-n\={n} states. 
One sees that the effect of isospin coupling, due to Coulomb and $\Delta_{np}$,  becomes negligible below few MeV
but it can account for  1 MeV difference nearthreshold, that is precisely in the region of Coulomb like protonium states.
The infinity of states, generated by Coulomb term (\ref{VC}) and with energies  of few KeV, are not represented in the figure.

\vspace{-0.5cm}

In summary, the models we have considered  contain  two parameters for each partial wave $\alpha$=$\{L,S,J\}$: the cut-off radius $r_c$ and the 
dimensionless annihilation  strength $\lambda$.
This gives two free parameters per uncoupled states ($^1$S$_0$,$^1$P1, $^3$P$_0$...). For the tensor coupled states, e.g. $^3$SD$_1$, this  means 2 parameters for $^3$S$_1$, 2 for $^3$D$_1$ plus
an additional cut-off parameters $r_T$  needed to regularise the tensor coupling potential $V_{^3S_1\to ^3D_1}$.

Next section will be devoted to present the scattering results.

\subsection{Protonium S-waves scattering lengths}

Our aim  is to fix the model parameters ($r_c,\lambda$) for the protonium S-waves   in order to reproduce the  experimental values of the singlet ($a_0$) and triplet ($a_1$) $p\bar{p}$ scattering lengths.
These values  are not directly measured but we have extracted them  from the measured \cite{Ausberger99}   protonium level shifts and widths  (\ref{DE})
by means of the Trueman expansion \cite{Trueman_NP26_1961}   in the way described in \cite{CRW_ZPA343_1992}.
The results for $\Delta E$ and $a_s$ are given in Table  \ref{Exp_a0_protonium} together with the predictions of the KW optical model obtained in  \cite{CIR_ZPA334_1989}.

\begin{table}[h!]
\begin{center}
\begin{small}
\begin{tabular}{l l ll }
                                    &Exp \cite{A_NPA658_99}                                   &    KW   \cite{CIR_ZPA334_1989,CRW_ZPA343_1992}                        \\\hline 
$\Delta E({^1S_0})$    &   0.440(75) - i 0.600(125)                                   &    0.50-i 0.63            \\
$\Delta E({^3SD_1})$  &  0.785(35) - i 0.470(40)                                     &    0.78-i 0.49         \\\hline
a$_0$($^1S_0   $)        &   0.49(9) - i 0.73(14)                                         &    0.57- i 0.77    &         \\
a$_1$($^3SD_1$)  &  0.93(4) - i 0.60(5)                                                  &    0.92- i0.63     &       \\
\end{tabular}
\end{small}
\end{center}
\caption{Experimental  S-wave protonium complex energy shifts $\Delta E$ (in keV) 
and  p\={p} scattering length $a_{s=0,1}$ (in fm)  together with the KW optical model predictions taken from \cite{CIR_ZPA334_1989,CRW_ZPA343_1992}.
The experimental $a_s$ values are extracted from the measured $\Delta E$ by means of the Trueman relation \cite{Trueman_NP26_1961}.}\label{Exp_a0_protonium}
\end{table}

For methodological reasons we have proceeded in two consecutive steps.

\begin{figure}[h!]
\begin{center}
\vspace{-0.3cm}
\includegraphics[width=9.cm]{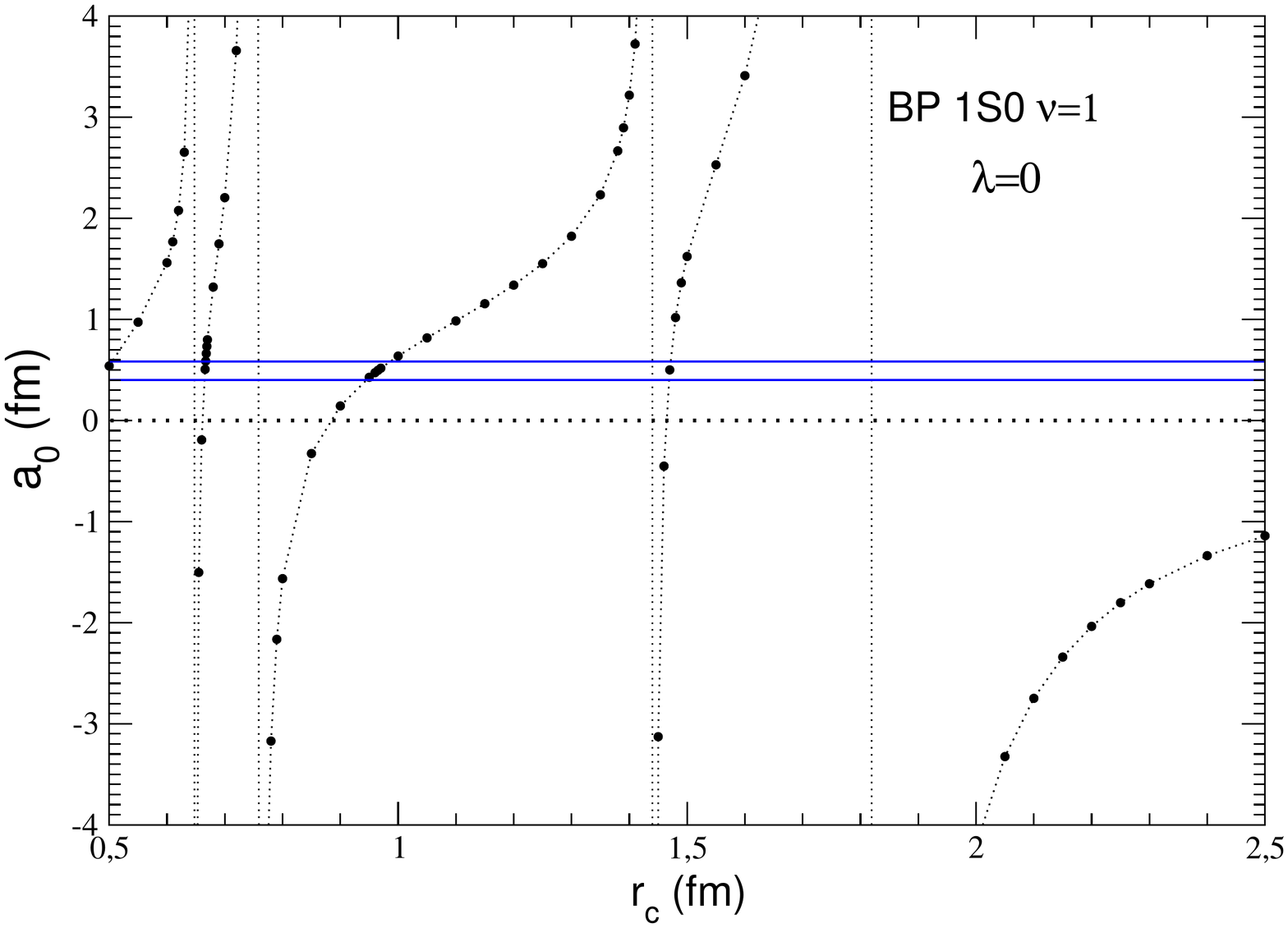} 
\vspace{-0.5cm}
\includegraphics[width=9.cm]{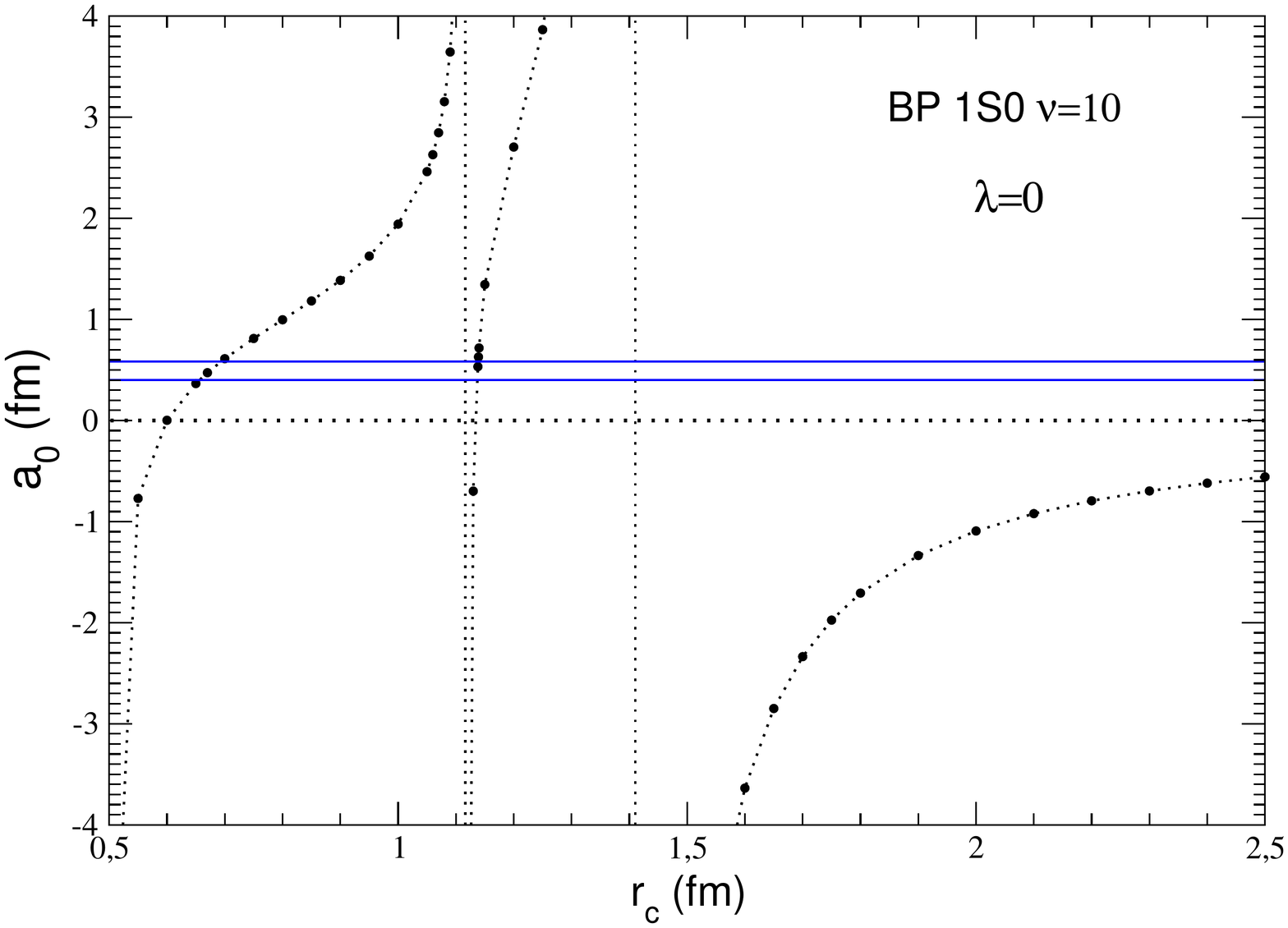}
\vspace{-0.2cm}
\caption{$^1$S$_0$ $\bar pp$ scattering   as a function of the $r_c$ without coupling to annilihation channels
with Model I regularization parameter  $\nu$=1 (upper panel) and $\nu$=10 (lower panel) in (\ref{VREG}) }\label{a0_rc_1S0_lambda_0}
\end{center}
\end{figure}

As a first step we have studied the dependence of the singled $^1$S$_0$ $\bar{p}p$ scattering length $a_0$ as a function of the cut-off radius $r_c$ in absence of any annihilation, i.e. $\lambda(^1S_0)=0$.
The results for Model I (BP) are displayed in Figure  \ref{a0_rc_1S0_lambda_0} for the two regularisation parameter values $\nu$=1 (upper panel) and $\nu$=10 (lower panel).
The blue horizontal lines correspond to the experimental value of Re[$a_0$]=0.49(9) fm.

One can see that for large values of $r_c$ the sign of $a_0$ is  negative, while
the experimental  value has $\mathrm{Re}(a)=0.49(9) >0$.
This negative sign corresponds to an attractive $p\bar{p}$ potential, obtained by averaging the isospin components of $V_{N\bar{N}}$  displayed in Figure   \ref{VNNB_1S0_rc}
plus the attractive Coulomb force.
When $r_c$ is decreased,  $V_{p\bar{p}}$  becomes increasingly attractive until the first $p\bar{p}$ bound state appears. 
This corresponds to a divergence of $a_0$ which happens  at $r_c\approx 1.8$ fm for $\nu$=1  (see also Figure \ref{B_rc_BP_1S0_n_1}) and  $r_c\approx 1.4$ for $\nu$=10.
 In order to obtain a positive scattering length we are obliged to accommodate one (or several) bound states.

\begin{figure}[h!]
\begin{center}
\vspace{-0.3cm}
\includegraphics[width=9.cm]{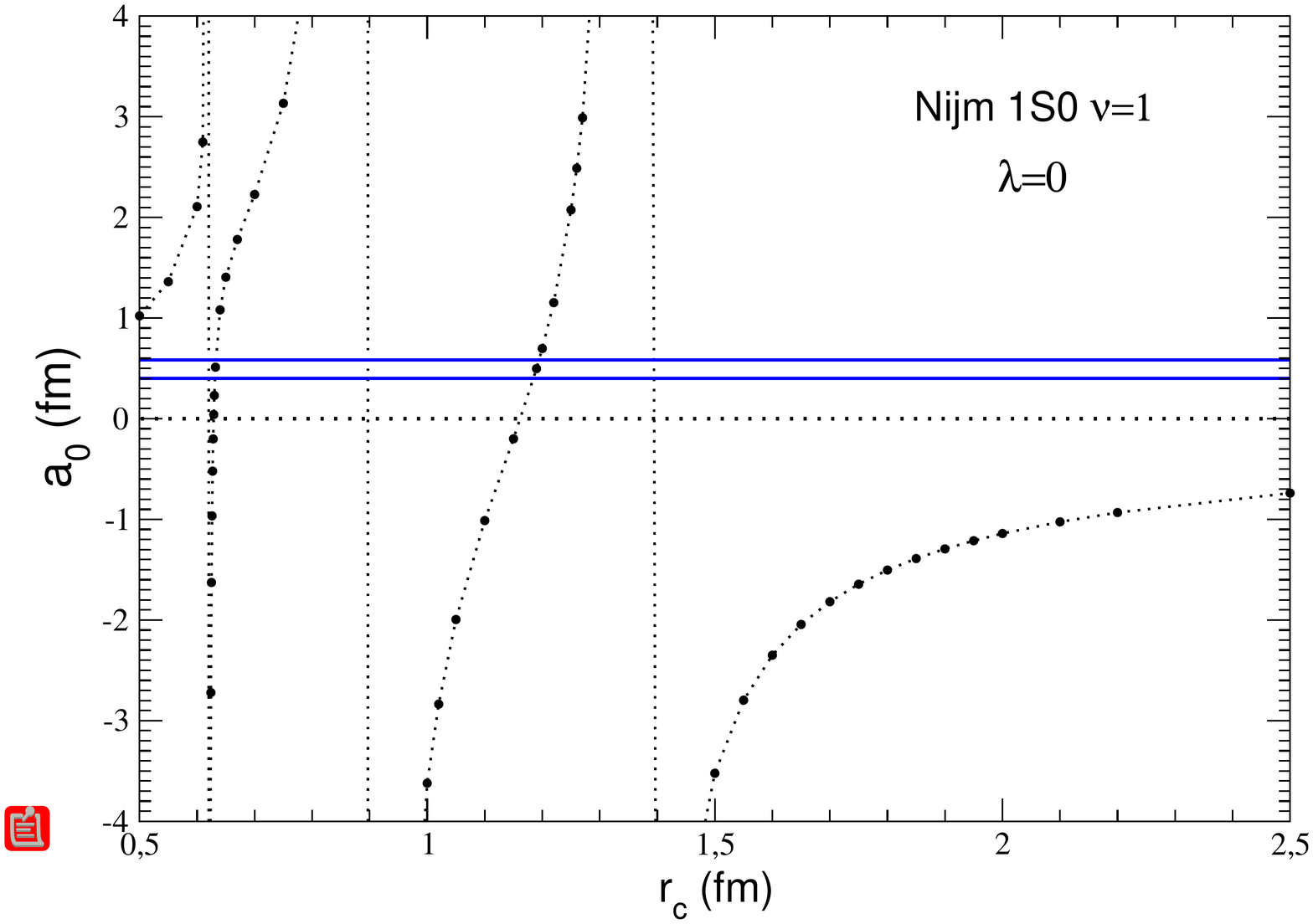}\vspace{-0.3cm}
\includegraphics[width=9.cm]{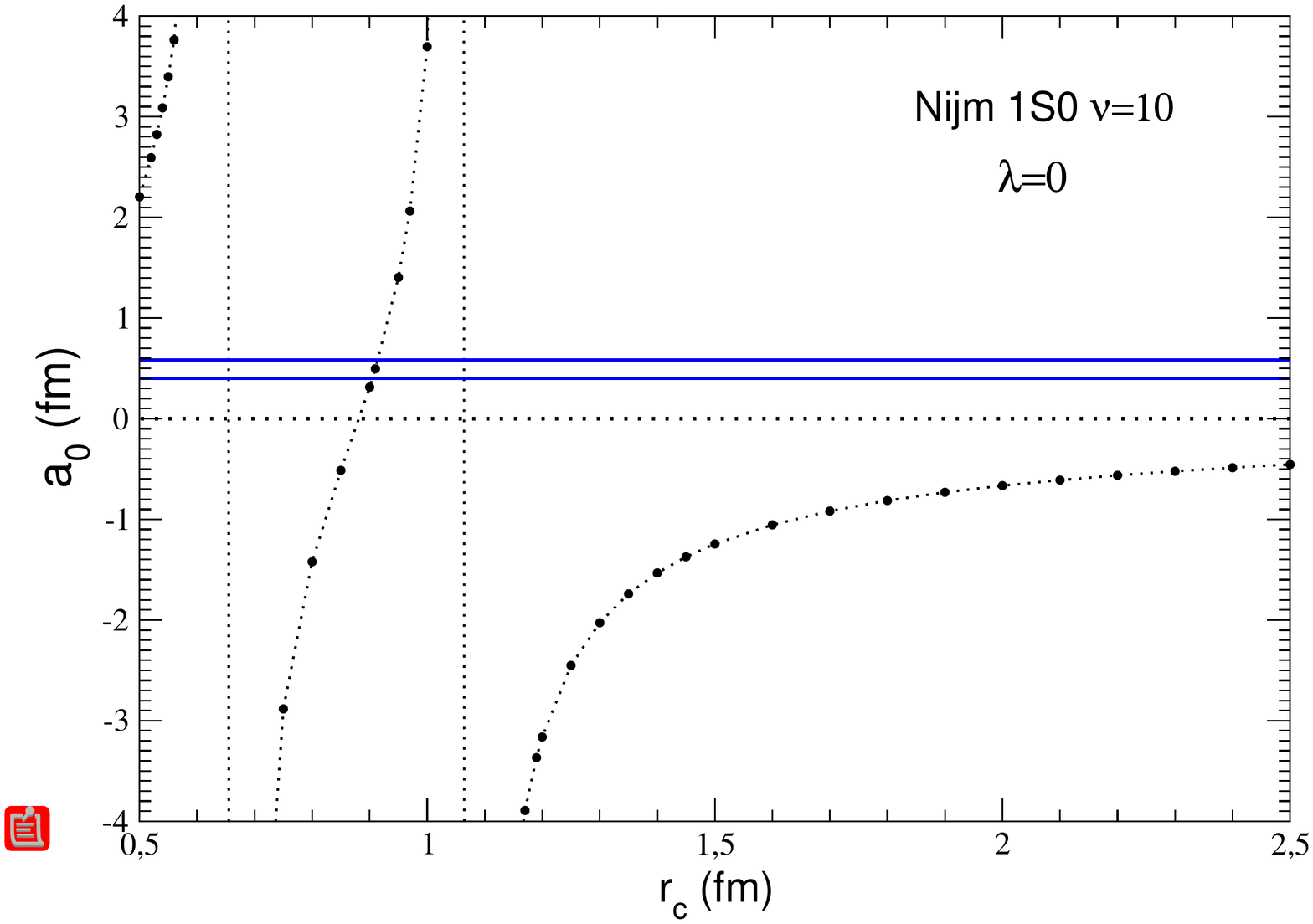}
\vspace{-0.3cm}
\caption{The same than Figure   \ref{VNNB_1S0_rc} for Model II, based on  Nijmegen meson exchange potential of \cite{NNB_CC_Nijm_1984}.}\label{a0_rc_1S0_lambda_0_Nijm}
\end{center}
\end{figure}

The same  conclusion is reached in Model II, when using the  N\={N} potential built with the Nijmegen meson contents of Ref. \cite{NNB_CC_Nijm_1984}.
The results are represented in Figure \ref{a0_rc_1S0_lambda_0_Nijm}, with $\nu$=1 in the upper part and $\nu$=10 in the lower one. 
Nijmegen potential is much weaker than BP  specially for the T=0 components -- due to the greater number of mesons
involved that  favours  large cancelations between their contributions -- but it  displays the same qualitative result with
the first singularity at $r_c\approx 1.4$ fm for $\nu$=1 and  $r_c\approx 1.1$ fm for $\nu$=10.

Thus, we would like to emphasize that the existence of  quasi-nuclear states (or baryonium) suggested by Dalkarov, Shapiro and Mandeltsveig 50 years ago \cite{DMS_JETPL10_1969,DMS_SJNP11_1970,DMS_NPB21_1970,DMS_PMNP_1971},
appears to be a natural, and  unavoidable, consequence of  the long range part of the NN interaction itself.

One could argue that a repulsive $V_{p\bar{p}}$ potential could also generate a p\={p}  Re[$a_{0}>0$] value without requiring any underlying bound state. 
However this is in contradiction  with  the bulk of  experimental data \cite{SH_PR35_1978}  as well as with all the existing \={N}N meson-exchange inspired models.
It would be very interesting to see whether a pionless EFT inspired \={N}N model, with low energy constants  directly fitted to experimental results, reach the same conclusion in what concerns the  \={N}N spectra.

In absence of any annihilation process we have at our  disposal several solutions to reproduce the experimental Re($a_0$): those given by the intersection
of the $a_0(r_c)$ curve with the two solid blue horizontal  lines representing the experimental value with the error band.
For Model I, based on BP potential, the solutions displayed in  Fig. \ref{a0_rc_1S0_lambda_0}
are: for $\nu$=1 $r_c^{(1)}\approx 1.47$  and $r_c^{(2)}\approx 0.96$,  while for $\nu$=10   $r_c^{(1)}\approx 1.10$ and $r_c^{(2)}\approx0.70$.
For Model 2, based on Nijmegen potential, the results are displayed in  Fig. \ref{a0_rc_1S0_lambda_0_Nijm}. 
For $\nu$=1 they give    $r_c^{(1)}\approx 1.2$ fm, $r_c^{(2)}\approx 0.6$ fm while  with $\nu$=10 one sees a unique solution with $r_c\approx 0.9$ fm.
Notice that other solution could exist for smaller values of $r_c$, and so involving increasingly larger potentials,  but we are rather interested in the more peripheral ones.
To this aim, Model I allows to use systematically larger cut-off  values.

\begin{figure}[!ht]
\begin{center}
\vspace{-0.3cm}
\includegraphics[width=9.cm]{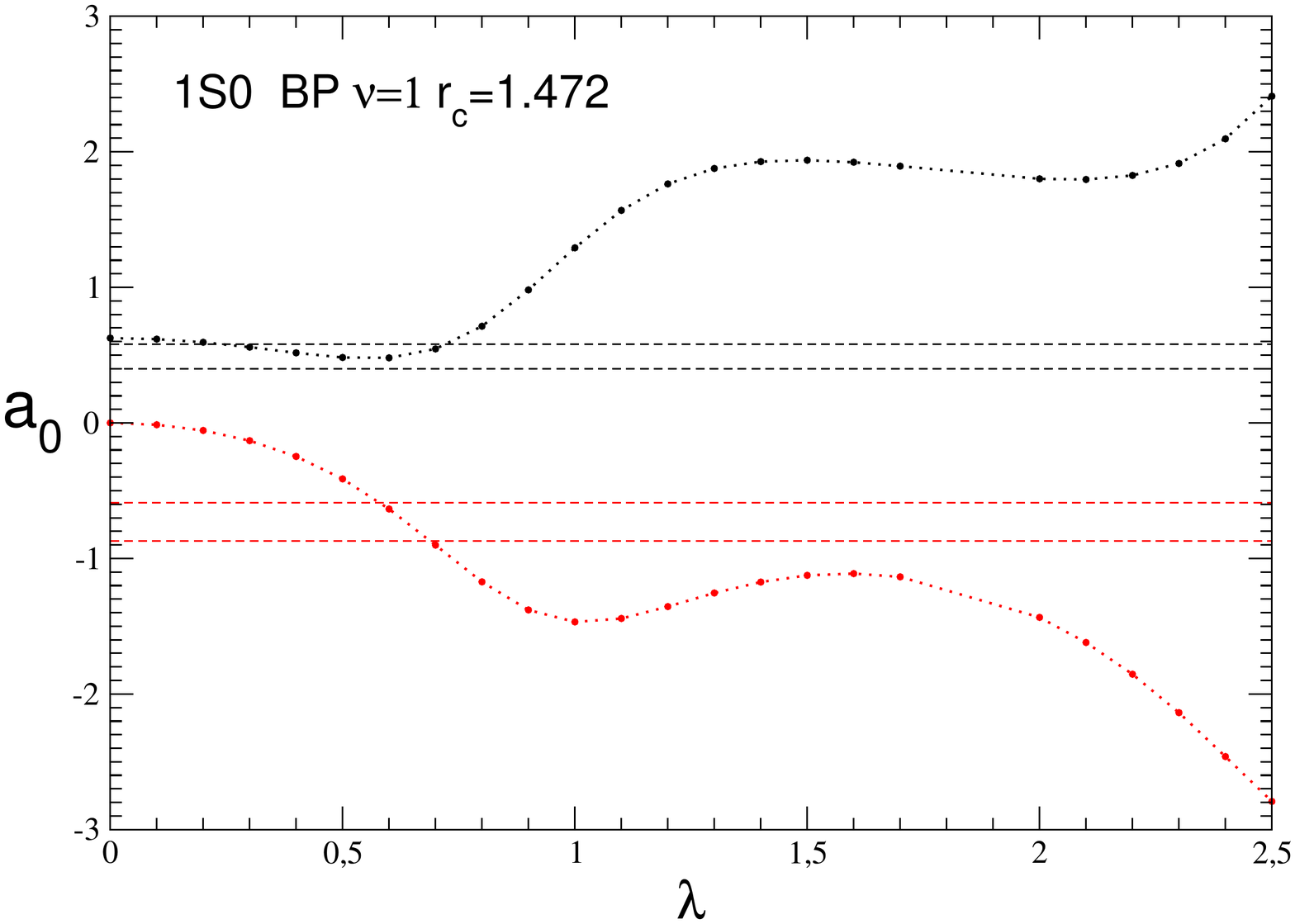}\vspace{-0.3cm}
\includegraphics[width=9.cm]{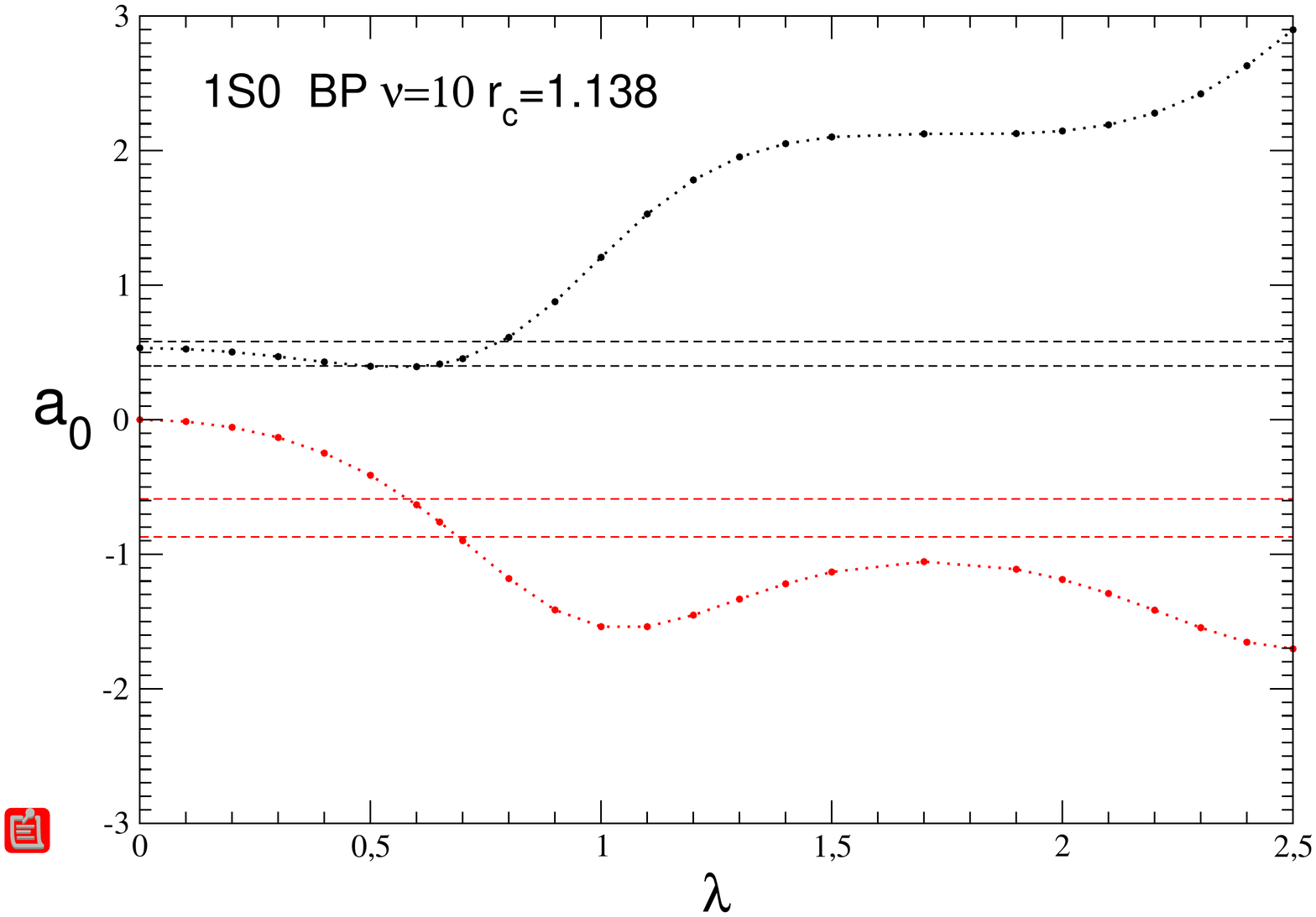}
\vspace{-0.3cm}
\caption{$^1$S$_0$   $\bar pp$ scattering   length (fm) as a function of the annihilation strength
$\lambda$ for Model I: upper panel corresponds to $\nu=$1 (with $r_c$=1.472 fm) and lower  to $\nu$=10 (with $r_c$=1.138 fm). 
Black solid lines indicate Re($a_0$) and red ones Im($a_0$). Horizontal dashed lines indicate the  corresponding experimental value from Table  \ref{Exp_a0_protonium}.}\label{a0_rc_1S0_lambda}
\end{center}
\end{figure}

The second step consists in  keeping fixed  the  $r_c$ values previously determined and  switching-on the annihilation potential.
The effect of the annihilation potential  modifies only slightly the $r_c$ values determined at $\lambda$=0.
For  Model I, we have taken the largest solutions of the $\lambda$=0 case, i.e. $r_c$=1.472 fm for $\nu=1$  and $r_c$=1.138 fm for $\nu$=10, 
and  we have increased the (dimensionless) annihilation strength $\lambda$ step by step.
The results are displayed in Figure \ref{a0_rc_1S0_lambda}, upper panel for $\nu$=1 and lower for $\nu$=10. 
Black solid lines indicate the real part of $a_0$ and red ones its imaginary part.
Dashed horizontal lines represents a band with the experimental values from Table \ref{Exp_a0_protonium} including errors, in black
for the real part and in red for the imaginary part.
As one can see the two values of $\nu$  give quite similar results and one obtains a satisfactory description  with $\lambda$=0.60  (for $\nu$=1) and with $\lambda$=0.65  for $\nu$=10.
Furthermore, this solution seems to be unique for a given $r_c$.

\begin{figure}[!ht]
\begin{center}
\vspace{-0.3cm}
\includegraphics[width=9.cm]{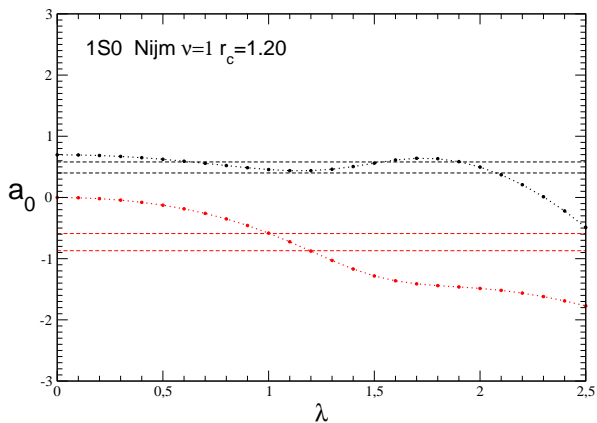}
\vspace{-.3cm}
\caption{Same as Figure  \ref{a0_rc_1S0_lambda}  but with Model II based on Njimegen potential and $\nu$=1.}\label{a0_rc_1S0_lambda_NIJM}
\end{center}
\end{figure}

The same results corresponding to  Nijmegen potential (Model II) are given in Figure  \ref{a0_rc_1S0_lambda_NIJM} .
The first $^1$S$_0$ solution  for $\nu=1$ is obtained  with the parameter values ($r_c$=1.20 fm, $\lambda$=1.10).

\vspace{-.cm}

A similar study has been performed for the  tensor coupled spin-triplet  state $^3$SD$_1$ in the two considered models.

\begin{figure}[!ht]
\begin{center}
\vspace{-0.3cm}
\includegraphics[width=9.cm]{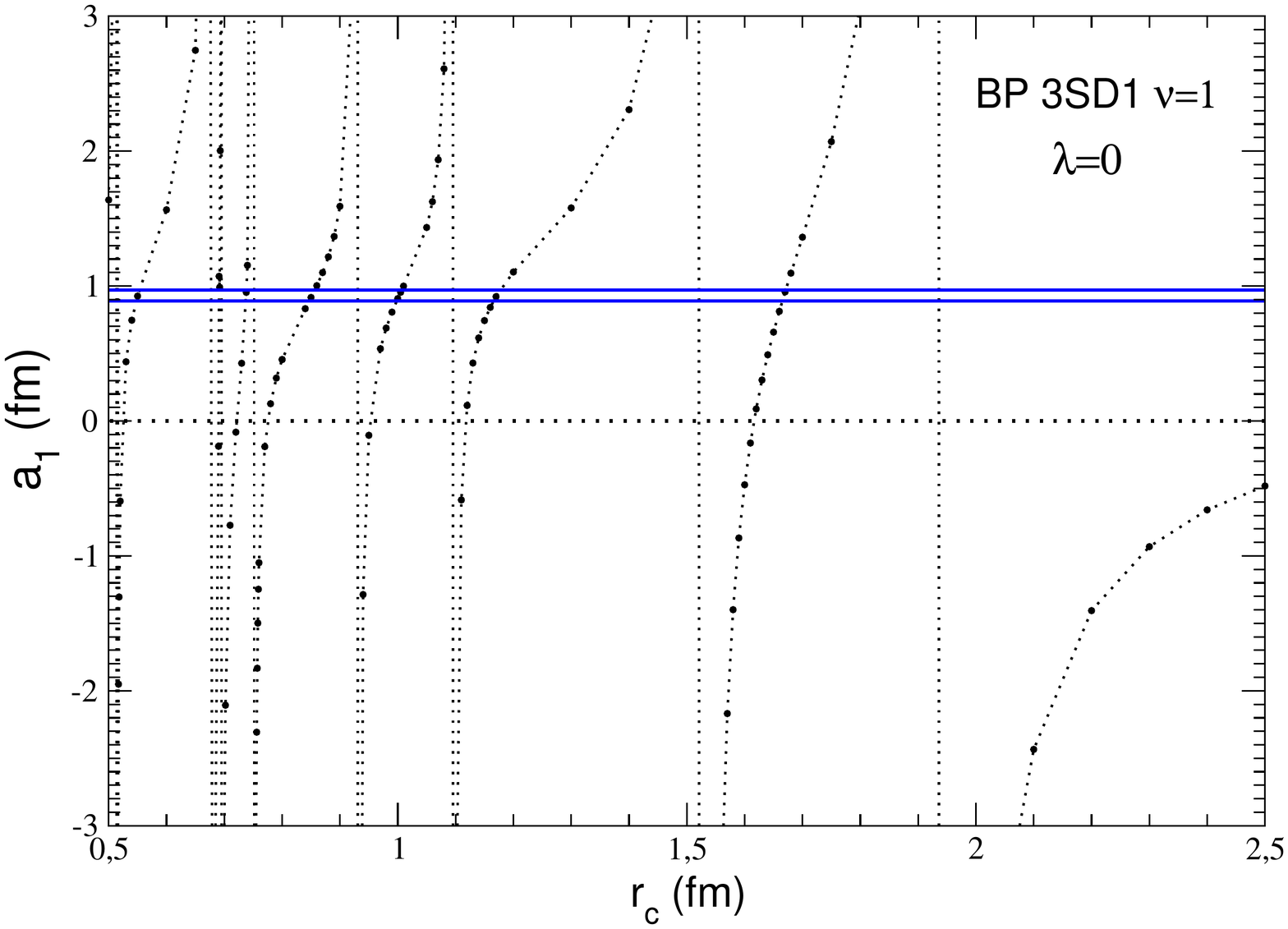}\vspace{-0.3cm}
\includegraphics[width=9.cm]{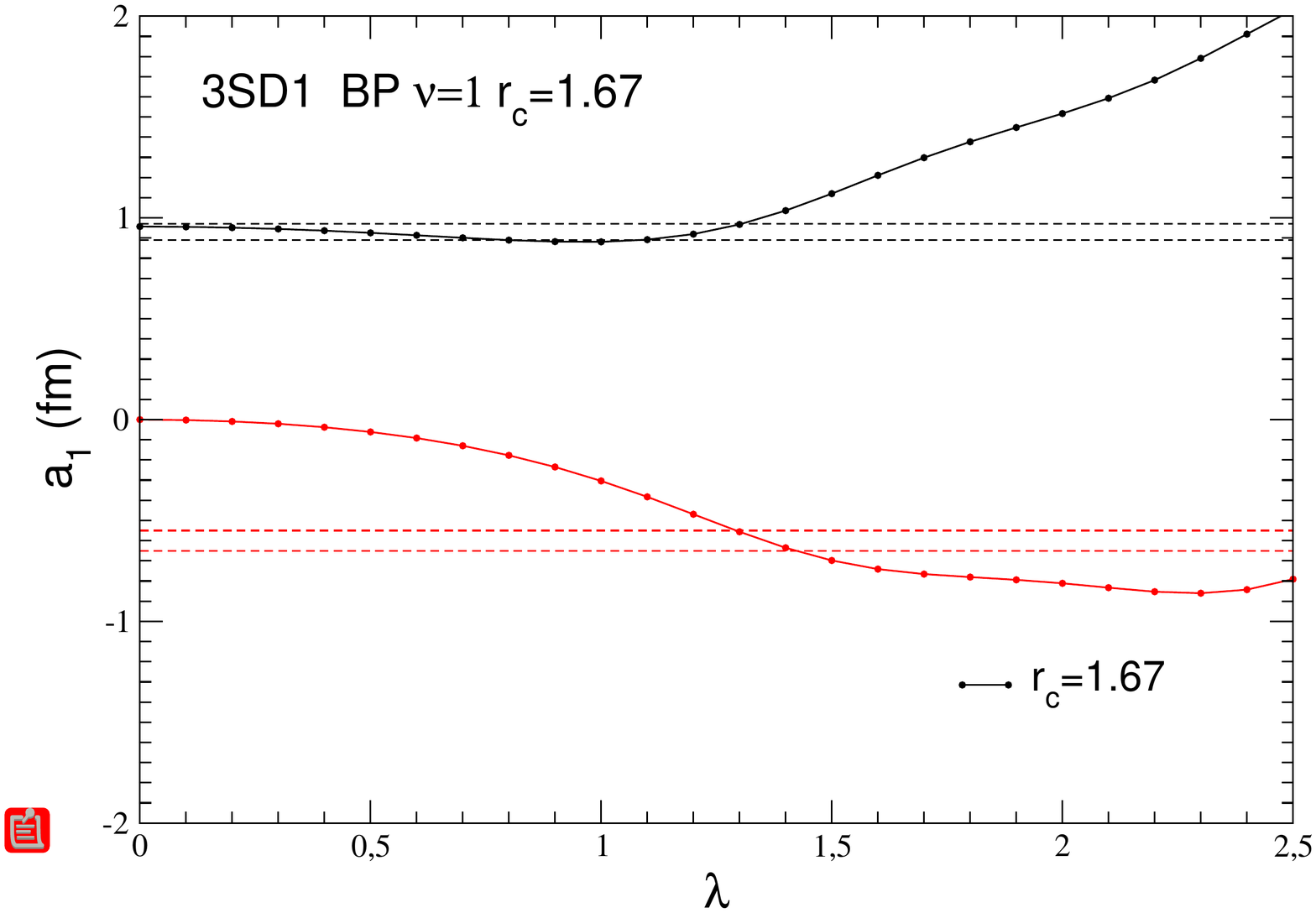}
\vspace{-0.3cm}
\caption{Upper panel : $^3$SD$_1$ $\bar pp$ scattering  length  as a function of  $r_c$ without coupling to annihillation channels
with Model I parameters  and $\nu$=1. Lower panel $\lambda$-dependence at fixed $r_c$=1.67 fm.}\label{a0_rc_3SD1_lambda_0}
\end{center}
\end{figure}

For BP potential (Model I), the $a_1$ dependence on the cut-off radius is shown in the upper panel of Figure \ref{a0_rc_3SD1_lambda_0}.
Again, for large  $r_c$ values the  scattering length is negative, contrary to the experimental value, and becomes
positive only after the the first bound state appears.  That happens at $r^{(1)}_p\approx 1.9$ for $\nu$=1.
Several other bound state exist at smaller values $r^{(2)}_p\approx 1.53$ fm, $r^{(3)}_p\approx 1.09$ fm ...
The qualitative behaviour of $a_1(r_c)$  is quite close to the singlet $^1$S$_0$ one (upper panel of Figure  \ref{a0_rc_1S0_lambda_0}). 
However, due to the strongness of the triplet N\={N} potentials, 
the $a_1(r_c)$ dependence displays a much richer structure of bound states and a larger number of solutions than for $^1$S$_0$.
By selecting the most  peripheral solution  compatible with the experimental Re[$a_1$]  ($r^{(1)}_c\approx 1.67$ fm for $\nu$=1) 
and switching  on the annihilation, one can reproduce the protonium $^3$SD$_1$ data with the parameters values ($r_c$=1.67 fm ,$\lambda=1.30$).
This is shown in the lower panel of  Figure \ref{a0_rc_3SD1_lambda_0}.

\begin{figure}[!ht]
\begin{center}
\vspace{-0.3cm}
\includegraphics[width=9.cm]{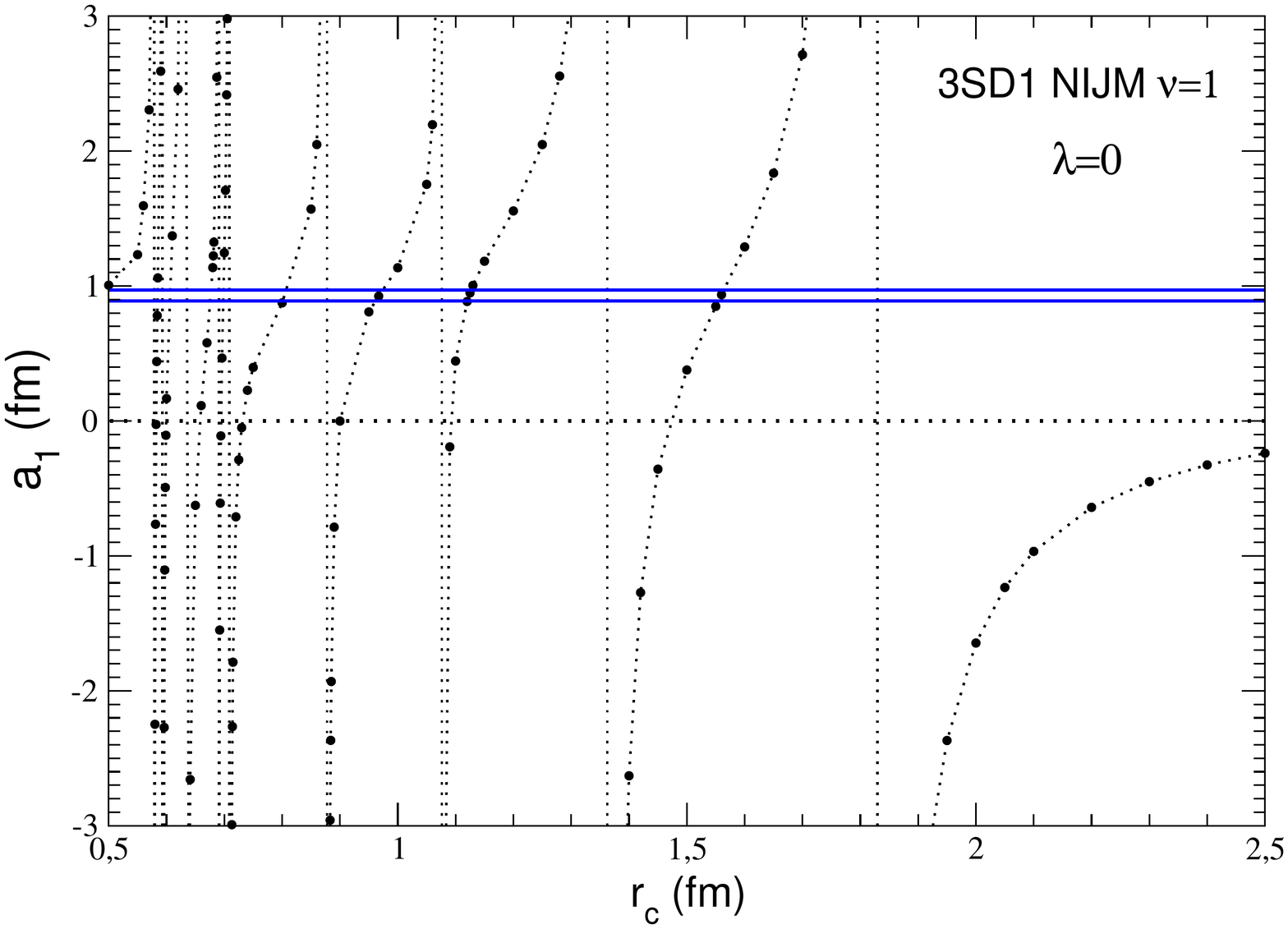}\vspace{-0.4cm}
\includegraphics[width=9.cm]{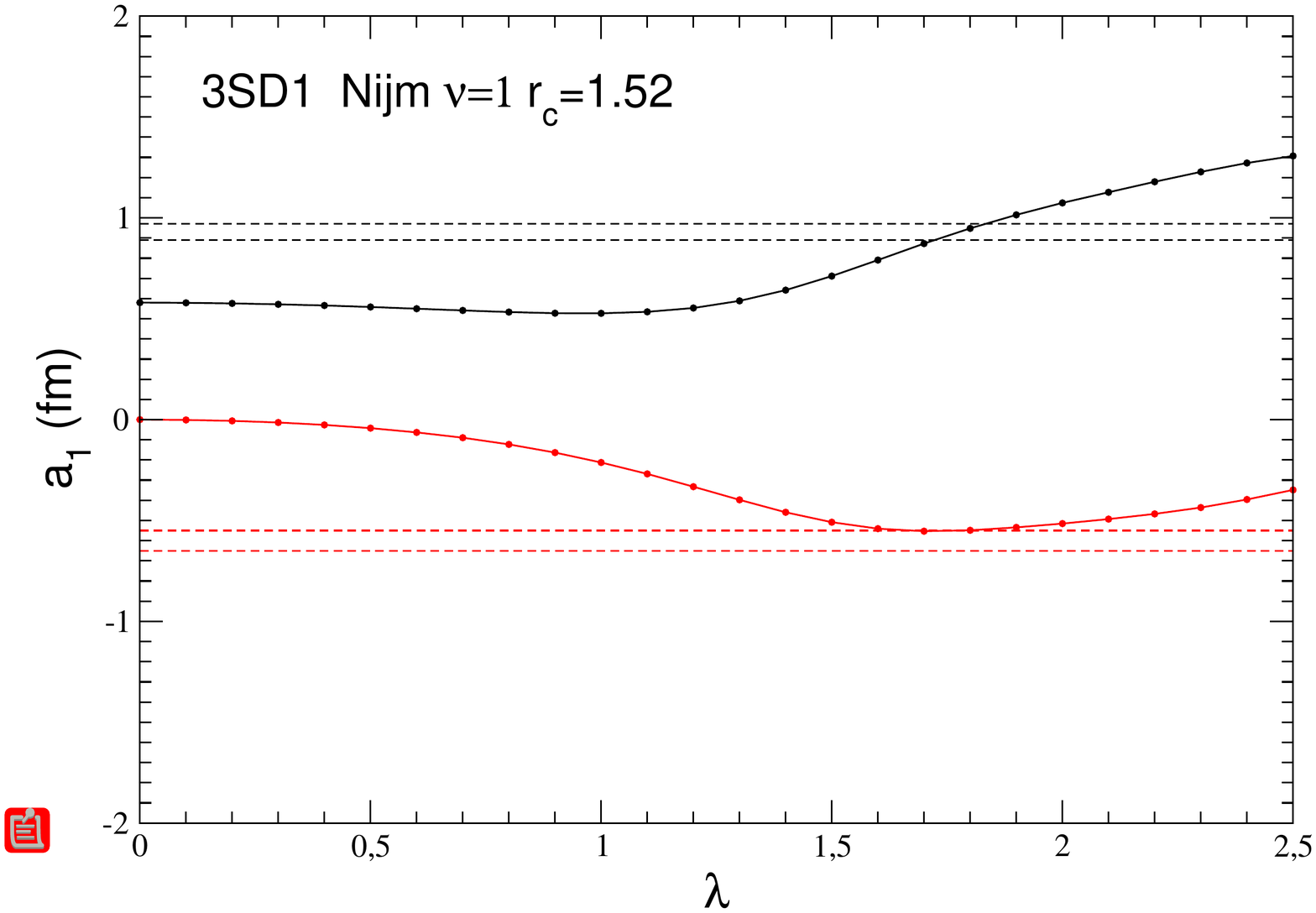}
\vspace{-0.4cm}
\caption{The same than figure \ref{a0_rc_3SD1_lambda_0} 
for Model II based on  Nijmegen potential. Upper panel : $^3$SD$_1$ $\bar pp$ scattering length  as a function of  $r_c$ without coupling to annihillation channels. 
Lower panel $\lambda$-dependence at fixed $r_c$=1.52 fm.}\label{a0_rc_3SD1_lambda_0_NIJM}
\end{center}
\end{figure}

By considering Model II, based on Nijmegen potential,  the same qualitative agreement is found.
We have displayed in Figure  \ref{a0_rc_3SD1_lambda_0_NIJM}  (upper panel)   the corresponding $a_1$ dependence on $r_c$ with $\lambda$=0 
and in the lower panel its  $\lambda$-dependence  for $r_c=1.52$ fm. This provides  a solution reproducing the experimental data  with ($r_c$=1.52 fm, $\lambda$=1.80).

All together, this allowed us to determine the S-wave model parameters, which are listed in Table  \ref{CC_parameters_a_0_protonium}.
Notice that the cut-off radii of Model I are systematically larger than those of Model II and the corresponding annihilation strengths $\lambda$ are smaller.

\begin{table}[h!]
\begin{center}
\begin{tabular}{l c c c c }
\multicolumn{5}{c}{Model I }\\\hline  
State                      &$\nu$     &  $r_c$(fm)   &  $\lambda$ &  $a_0$ (fm)\\\hline 
$^1S_0$                & 1    &    1.47  &  0.60    &   0.495 -i 0.747  \\
                              &       &    0.93    &  1.91    &   0.498 -i 0.716     	 \\
                              & 10  &    1.14  &  0.65    &   0.458 -i 0.770     	 \\ \hline
$^3SD_1$             & 1    &    1.67  &  1.30    &   0.936 -i 0.556    \\
                              & 10  &    1.31  &  1.55    &   0.934 -i 0.608    \\
\end{tabular}

\bigskip
\begin{tabular}{l c c c c }
\multicolumn{5}{c}{Model II }\\\hline  
State                      &$\nu$     &  $r_c$(fm)  &  $\lambda$ &  $a_0$ (fm)\\\hline 
$^1S_0$                & 1    &    1.20        &  1.10    &   0.438 - i 0.724  \\ 
                             & 10   &    0.91       &   1.40    &   0.446  - i 0.785  	 \\ \hline
$^3SD_1$             & 1    &    1.52      &  1.75    &    0.948  - i 0.549        \\
                              & 10  &    1.18      &  2.00    &     0.800 - i 0.480 \\
\end{tabular}
\end{center}
\caption{Computed  p\={p} scattering lengths and corresponding parameters with  Models I and II}\label{CC_parameters_a_0_protonium}
\end{table}

Notice also  that although the solutions allow relatively large cut-off  values, the role  of the OBE contributions in the resulting \={N}N potential is essential. 
Indeed, at $r\approx 1.4$ fm, the strong N\={N} potential  displayed in Figure  \ref{VNNB_1S0_rc} is $\sim$ 30  MeV, while the Coulomb p\={p}  potential is $\sim$ 1 MeV.
Even at such distances the N\={N}  interaction is dominated by one-boson exchanges, mainly pion but still sizeable contributions from heavier mesons.

\vspace{-.cm}

\subsection{Protonium level shifts and annihilation densities}

Once the potential parameters have been fixed by reproducing the experimental $p\bar{p}$ scattering lengths,
the four-channel homogeneous Schrodinger equation (\ref{SCHEQ}) for protonium bound states was solved in coordinate space, using a spline decomposition. 

Thus, using the parameters of Mode I  ($\nu=1$)  displayed in Table \ref{CC_parameters_a_0_protonium} 
we have obtained  for  $^1$S$_0$ state,  
a level shift   $\epsilon_R$=0.427 keV and $\Gamma$= 1.300 keV  ($r_c$=1.472, $\lambda$=0.65) and
a level shift $\epsilon_R$=0.426 keV and $\Gamma$=1.278 keV  ($r_c$=0.93, $\lambda$=1.91).
As expected, these protonium  $\Delta E$'s are in good agreement with the experimental data  from  Table  \ref{Exp_a0_protonium}.
In addition we have at our disposal the corresponding realistic S-wave protonium wavefunctions in the framework of a UCCM
to obtain the corresponding  annihilation densities.

It is worth noticing here that, despite UCCM gives similar prediction than optical models --  both are adjusted to this aim ! --  there are substantial differences 
in their underlying dynamics and the short range  wave functions.

\begin{figure}[h!]
\begin{center} 
\vspace{-0.3cm}
\includegraphics[width=9.cm]{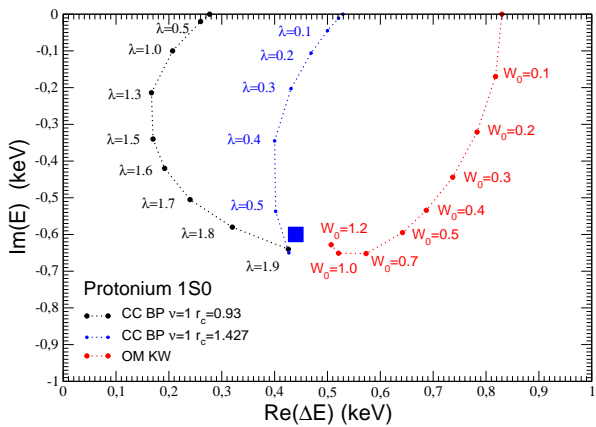}
\vspace{-0.5cm}
\caption{Evolution of the $^1$S$_0$  complex level shifts $\Delta E$ as a function of  the annihilation strength 
for UCCM (dashed black and blue lines) and  OM (red dashed lines).} \label{E_lambda_1S0_rc_093}
\end{center} 
\end{figure}

The first one  concerns the dependence of the complex energies $E$ on the annihilation strength.
This was already anticipated at  the end of  section \ref{UCC_Models} for a deeply bound quasi-nuclear state and we will now examine the differences
in a Coulomb-like, loosely bound, protonium state.
To this aim we have displayed in Fig. \ref{E_lambda_1S0_rc_093} the trajectory of $\Delta E$  
in the complex energy plane for the $^1S_0$ state as described by  the UCCM (black and blue symbols, corresponding to two different solutions $r_c=$1.472 fm and $r_c$=0.93 fm)  and the Khono-Weise OM (red symbols). 
These trajectories are parametrised by  the annihilation strength:  $W_0$ in OM and $\lambda$ in UCCM.
Both of them end up close to the experimental point (denoted by a filled blue square) and are compatible within errors.
The maximal annihilation strength is $W_0$=1.2 GeV for OM, and  $\lambda=0.65$ and $\lambda$=1.91 for UCCM.

Its behaviour is not trivial,  when compared to the deeply bound state of Figure  \ref{Pole_Traj_11S0_QN}. 
For moderate values of $\lambda$ or $W_0$,   Im(E) is in both cases monotonously increasing.  
However, while for UCCM Im(E)($\lambda$) remains  increasing,  in OM the width of the state saturates at $W_0\approx 0.7$ GeV  and start to decrease for $W_0>1$ GeV.  
The real part of $\Delta E$ (level shifts $\epsilon_R$)  evolves in opposite directions than for the deeply bound state. 
For OM the effect of annihilation becomes attractive, i.e. decreasing $\epsilon_R$, for all values of $W_0$, while
in the UCCM the attraction is  limited to moderate value of $\lambda$ and becomes repulsive only close to the maximal values of 
$\lambda$ ($\lambda\approx$ 0.5 for $r_c$=1.427 fm and $\lambda\approx$ 1.8 for $r_c$=0.93 fm).
The difference with respect to the deeply bound state are due to the structure of the wave function. 
In the former case the size of the state is $\sim$ 1 fm, a spatial region where the annihilation process takes place,
while for the protonium states its spatial dimension is given by the Bohr radius $B\approx$ 57 fm. 

\begin{figure}[h!]
\begin{center} 
\vspace{-0.3cm}
\includegraphics[width=9.cm]{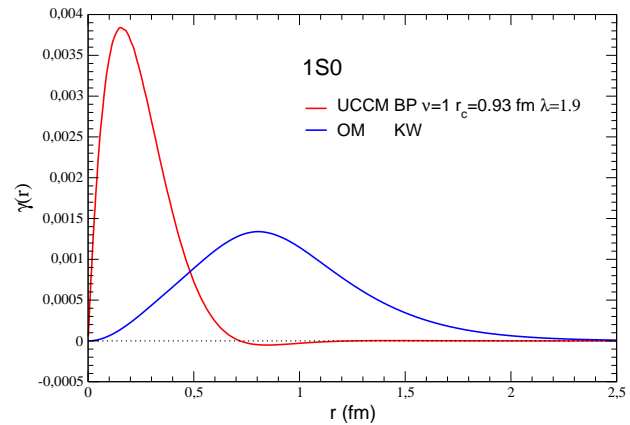}
\end{center}
\vspace{-0.5cm}
\caption{Protonium  annihilation density for the $^1$S$_0$ state described with the UCCM (in red) and with OM (in blue).
Both models reproduce the same experimental complex level shift $\Delta E$ value of Table \ref{Exp_a0_protonium}.}\label{Fig_PAD_1S0}
\end{figure}

The second one concerns the annihilation densities $\gamma$.
We have plotted in  Figure  \ref{Fig_PAD_1S0} the function  $\gamma(r)$ of the same protonium $^1$S$_0$ state 
described with the  UCC-Model I  (in red) and  the Kohno-Weise optical model (in blue). Both models reproduce equally well the experimental values of Table \ref{Exp_a0_protonium},
in particular the total width of the state $\Gamma$, i.e. the surface below $\gamma$.
However the annihilation process takes place in quite different ways: in the UCCM it is peaked around $r\approx0.15$ fm, given by the annihilation range $r_a$ of Figure \ref{NNB_annihil},
and spread over $\Delta r\approx0.3$ fm,
in OM  is peaked around $r\approx0.9$ fm and has a larger dispersion $\Delta r\approx1$ fm.
In the latter case it is difficult to relate the spatial distribution of $\gamma$ with the parameters of the annihilation potential $W(r)$ given in Table \ref{Tab_Par_OM}, 
since different choices  produce quite similar  results for $\gamma$  \cite{CIR_ZPA334_1989}.  

These differences are due to the way in which the annihilation  is modelled, as it was  described in previous sections \ref{O_Models} and \ref{UCC_Models}. 
Although describing the same total width $\Gamma$, they introduce a strong model dependence in the way we understand this process, in particular  its spatial distribution.
In one case (UCCM), annihilation is a short range ($r_a$) process well localised in the \={p}p inter-particle distances  while in the OM it is spread over a sizeable spatial region.
The origin of these differences is also due to the striking differences in the short range part of the \={p}p wave function.

\begin{figure}[h!]
\begin{center} 
\vspace{-0.3cm}
\includegraphics[width=9.cm]{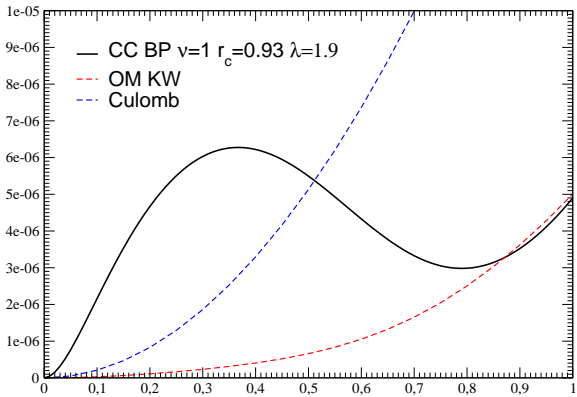}
\vspace{-0.3cm}
\caption{Modulus squared of the 1S0 $p\bar{p}$ radial wave function  in UCCM (solid black line) compared with the KW optical model  (in dashed red) and the pure Coulomb results (blue dashed line. \label{Fig:wf}}
\end{center}
\end{figure}

It is indeed interesting to compare  the  $p\bar{p}$ wave functions  provided by the UCCM   with the results 
of the optical model  and with the pure Coulomb wave function. 
This is done in Fig.~\ref{Fig:wf} where we have displayed the modulus squared of the reduced radial wave function $u(r)$ for the protonium $^1$S$_0$ state (first channel).
As one can see the results of the UCCM (black solid line) differ considerably from those of the  KW  optical model (dashed red line), in particular for small values of $r$.
In fact for $r\le 0.5$ fm, where the annihilation process take place, the OM wave function is significantly damped with respect to both the UCCM and the Coulomb (dashed blue line). 
At $r=0.15$ fm e.g. the CC wave function is two orders of magnitude  smaller than the OM one and it is even larger
than the Coulomb one despite the annihilation process.
As it was already mentioned, this striking differences are a direct consequence of the unitarity.

\section{Conclusion}\label{Conclusion}

We have considered  a unitary coupled channel model to describe the experimentally measured S-wave proton-antiproton 
scattering  length and the corresponding  hydrogen-like singlet and triplet protonium states. 

The  N\={N} interaction is obtained from a G-parity transform of two different  meson-exchange inspired NN models, regularized
at short distance.
We have shown that, in this theoretical framework, the existence of deep quasi-bound states of nuclear nature is an  unavoidable consequence of the protonium experimental data. 

The properties of these two types of states --  quasi-nuclear and hydrogenic  --   have been described 
and the differences with respect to the optical models description have been analyzed.

The  protonium  annihilation densities have been computed. When compared with the optical model results,  they show striking model differences in 
our theoretical understanding of  annihilation.
In particular with the unitary model, the annihilation  process is sharply peaked around 0.15 fm  while in optical models it is centered at $\approx 0.9$ fm and spread over one fm.

The differences in the short range part of the  proton-antiproton wave function have been examined.
In the case of optical models the wave function is strongly supressed due to the imaginary potential
while in the unitary coupled channel it is  two order of magnitude larger and it is even enhanced with respect to the Coulomb case.

\section*{Acknowledgement}
This work was performed during a research visit of EY supported by grant No.~2018/21758-2 from FAPESP. EY is thankful for the great hospitality shown to him during this visit to IJCLab.

\appendix

\section{Annihilation densities in  UCCM}\label{App_PAD}

Since the second channel is exponentially increasing, a direct application of (\ref{IE_OM})  cannot be applied.
However if in the expression
\begin{eqnarray}
E &=& E_R-i{\Gamma\over2}=  {1\over N^2} < \Psi \mid H \mid \Psi>  =  {1\over N^2} \cr
   &\times&   \int_0^R   \left[   u^*_1 H_{11}u_1+   u^*_2 H_{22}u_2  + V_{12}  {\rm  Re}(u_1^*u_2)   \right]  \label{IE_CC}   
\end{eqnarray}
with
\[   N^2=  \int_0^R dr \;  ( \mid u_1\mid^2 + \mid u_2\mid^2) \]
one restricts the integral at large but finit distance $R$, both -- numerator and denominator --
have the same kind of divergence and their ratio gives accurately the complex energy of the protonium state, which results
from the solution of the eigenvalue problem (\ref{SCHEQ_T}) with the appropriate boundary conditions.

Since the third term in (\ref{IE_CC}) is real, one has in particular
\begin{equation}\label{GammaT_CC_1}
{ \Gamma \over2}=  -  {\rm Im} \left\{ \int_0^R dr\; u^*_1 H_{11}u_1+  \int_0^R dr\; u^*_2 H_{22}u_2 \right\}  
\end{equation}
where we assume hereafter the use of -- finite distance -- normalized wave functions $u_c$.

By expliciting H and removing the real contributions
\begin{eqnarray*}
u^*_1 H_{11}u_1  &=& {\hbar^2\over2\mu_1}  \left\{  - u^*_1 u^{''}_1 +   \left[ {l_1(l_1+1)\over r^2}  + V_{11} \right] | u_1|^2  \right\}  \cr
u^*_2 H_{22}u_1  &=& {\hbar^2\over2\mu_2}  \left\{  - u^*_2 u^{''}_2 +  \left[ {l_2(l_2+1)\over r^2}    - \Delta\;\; \right] |u_2|^2   \right\}
\end{eqnarray*}
eq.  (\ref{GammaT_CC_1}) takes the form
\begin{equation}\label{GammaT_CC_2}
{ \Gamma\over2} =      \int_0^R dr  \; \left\{ {\hbar^2\over2\mu_1}\; {\rm Im} [ u^*_1 u_1{''}]   +  {\hbar^2\over2\mu_2} \; {\rm Im} [ u^*_2 u_2{''} ] \right\} 
 \end{equation}

However, we will see that the contribution of  exponentially decreasing solutions, like channel 1,  vanishes in the limit $R\to \infty$ i.e. :
\begin{equation}\label{Gamma1_CC}
  { \Gamma_1\over2} \equiv      \int_0^R dr  \;  {\hbar^2\over2\mu_1}\; {\rm Im} [ u^*_1 u_1'']  =  0 
 \end{equation}  
Indeed, let us integrate by parts the following expression, with $u_1$  regular at the origin:
\begin{small}
\begin{equation}\label{Gamma_1}
  \int_0^R  dr\;  u^*_1(r) u_1{''}r) =   u_1^*(R) u'_1(R) - \int_0^R dr  \mid u'_1(r)\mid^2   
\end{equation}
\end{small}
The second (integral) term is real and does not contribute to (\ref{Gamma1_CC}). The first term is given by the asymptotic behaviour  (\ref{ASYMP})
with the proper determinations for the channel momenta (\ref{q1}) and (\ref{q2}). Once Inserted in (\ref{Gamma1_CC}) it gives  
\begin{equation}\label{Gammap_1}
 {\Gamma_1\over 2}  =    {\hbar^2\over2\mu_1}    \beta_1 \mid N_1\mid^2 e^{ -2\beta_1 R }  \quad\rightarrow\quad 0
   \end{equation}
where $N_1$ is the asymptotic norm.
The first term in the r.h.s of eq. (\ref{GammaT_CC_2}) vanish and the total width $\Gamma$ of the state is given by eq. (\ref{GammaT_CC}).


\begin{thebibliography}{99}
\bibitem{Amsler_ANP18_87}  			C. Amsler, Adv. in Nucl. Phys. Vol 18, Ed. by J. Negele and E. Vogt, Plenum Press New York (1987) 183 
\bibitem{Gastaldi_NPA478_1988}  		U. Gastakdi, Nucl. Phis. A478 (1988) 813c

\bibitem{PUMA_LI_2017}   A. Obertelli, S. Naimi, T. Uesaka, A. Corsi, E.C. Pollacco, F. Flavigny, PUMA Letter of Intent, SPSC-I-247 (2017)
\bibitem{PUMA_CERN_Aproval_2021}  https://home.cern/news/news/physics/cern-approves-two-new-experiments-transport-antimatter

\bibitem{A_NPA658_99}  		M. Augsburger et al,        Nucl. Phys. A658 (1999) 149
\bibitem{REV_LEAR_KBMR_PREP368_2002}		E. Klempt, F. Bradamante, A. Martin, J.M. Richard, Phys Rep 368  (2002) 119
\bibitem{REV_LEAR_KBR_PREP413_2005}		E. Klempt, Ch. Batty, J.M Richard Phys Rep 413 (2005) 197-317
\bibitem{REV_NNB_Frontiers_2020} 			J.M. Richard, Front. Phys. 8 (2020) 6 


\bibitem{Ausberger99} M.~Ausberger et al, Nucl.~Phys.~A {\bf 658} (1999) 149. 

\bibitem{DR1_PRC21_1980}             		C.B. Dover and J.M. Richard, Phys. Rev. C21 (1980) 1466.
\bibitem{DR2_RS_PLB110_1982}          	J.M. Richard, M.E. Sainio,  Phys. Lett. B 110, 349 (1982)
\bibitem{KW_NPA454_1985}                 	M. Kohno, W. Weise, Nucl. Phys. A454, 429 (1986)


\bibitem{CLLMV_PRL_82}  			J. Cote, M. Lacombe, B. Loiseau, B. Moussallam, R. Vinh Mau, Phys. Rev. Lett. 48, 1319 (1982)
\bibitem{PLLV_PRC50_1994}   			M. Pignone, M. Lacombe, B. Loiseau, and R. Vinh Mau, Phys.  Rev. C 50, 2710 (1994). 
\bibitem{ELLV_PRC59_1999} 			B. El-Bennich, M. Lacombe, B. Loiseau, and R. Vinh Mau, Phys. Rev. C 59, 2313 (1999) 
\bibitem{PARIS_PRC79_2009}  		B. El-Bennich, M. Lacombe, B. Loiseau and S. Wycech, Phys. Rev. C79, 054001 (2009)

\bibitem{NNB_CC_Nijm_1984}		P.H. Timmers, W.A. van der Sanden, J.J. de Swart, Phys. Rev D 29 (1984) 1928

\bibitem{Haidenbauer_JHEP_2014}  		Xian-Wei Kang, Johann Haidenbauer and Ulf-G. Mei{\ss}ner,  JHEP 02 (2014) 113 
\bibitem{Haidenbauer_JHEP_2017}		Ling-Yun Dai, Johann Haidenbauer, Ulf-G. Mei{\ss}ner , JHEP 1707 (2017) 078

\bibitem{CIR_ZPA334_1989}   			J. Carbonell, G. Ihle, J.M. Richard, Z. Phys. A 334 (1989) 329-341

\bibitem{Bachir_PhD_Paris_1980}  	B. Moussallam, PhD Thesis, Universite Paris VI, 1980 (unpublished).

\bibitem{KKMS_JETPL26_1977}       B.O. Kerbikov, A.E. Kudryavtsev, V.E. Markushin and I.S. Shapiro, JETP Letters 26 (1977) 368  
\bibitem{SH_PR35_1978}			I. S. Shapiro, Phys. Rep. 35C (1978)129 
\bibitem{CDPS_NPA535_91}		J. Carbonell, O.D. Dalkarov, K.V. Protasov, I.S. Shapiro, Nucl. Phys. A535(1991)651-668

\bibitem{Gasiorowicz_1966}    Stephen Gasiorowicz., Elementary Particle Physics. Wiley, New York, 1966

\bibitem{MARTIN 61}A. Martin, Phys. Rev. 4 (1961) 614

\bibitem{BMS_SJNP30_1979} 		L.N. Bogdanova, V.E. Markushin and I.S. Shapiro, Sov. J. Nucl. Phys. 30(2)  (1979)  248; Yad. Fiz. 30, 480-496 (1979) 

\bibitem{DMS_JETPL10_1969}   	O. D. Dalkarov, V. B. Mandelzweig, and I. S. Shapiro, Zh. Eksp. Tear. Fiz., Pis'ma Red. 10, 402 (1969) [JETP Lett. 10, 257 (1969)].
\bibitem{DMS_SJNP11_1970}   	O.D. Dalkarov, V. B. Mandelzweig, and I. S. Shapiro, Yad. Fiz. 11, 889 (1970) [Sov. J. NucL Phys. 11, 496 (1970)].
\bibitem{DMS_NPB21_1970}	 	O.D. Dalkarov, V.B. Mandelzweig, and I. S. Shapiro, Nucl. Phys. B 21, 88 (1970
\bibitem{DMS_PMNP_1971}		O.D. Dalkarov, V. B. Mandelzweig, and I. S. Shapiro, Problemy sovremennol yadernol fiziki (Problems of Modern Nuclear Physics), Nauka, 1971, p. 132.

\bibitem{DPS_IJMPA5_1990}  	O. D. Dalkarov, K. V. Protasov, I. S. Shapiro, Int. J. Mod. Phys. A5 (1990)1995
\bibitem{DCP_SJNP52_1990}	O. D. Dalkarov, J. Carbonell, K. V. Protasov, Sov. J. Nucl. Phys. 52, 6(1990)1052
\bibitem{CPD_NP558_1993} 	J. Carbonell, K. V. Protasov, O. Dalkarov, Nucl. Phys. A 558 (1993) 353
\bibitem{CPD_PLB306_1993}	J. Carbonell, K. V. Protasov, O. Dalkarov, 	Phys. Lett. B 306 (1993) 407
\bibitem{DCK_NCA107_1994} 	O.D. Dalkarov, J. Carbonell, K.V. Protasov, Nuovo Cim. A107 (1994) 2409


\bibitem{BP_NPB5_1968}    	R. A. Bryan, R. J. N. Phillips, Nucl. Phys.{\bf B5} (1968) 201

\bibitem{Trueman_NP26_1961}         T.L. Trueman,  Nucl. Phys. 26, 57 (1961)
\bibitem{CRW_ZPA343_1992}      	J.Carbonell, J.M. Richard and S.Wycech, Z.Phys. A343 (1992) 325.

\bibitem{Erice_1988} I.S. Shapiro  Lectures at the International School  Antiproton-Nucleon and Antiproton-Nucleus Interactions,  
          Ettore Majorana Center for Scientific Culture, Erice 10-18 june (1988),  
          Editors: Bradamante, F., Richard, J.M., Klapisch, R. (Eds.), Plenum Press New-York  (1990) pag 81
\bibitem{Mainz_1988}   European Symposium  ''Proton anti-proton interactions and fundamental symmetries'', Mainz,  September 5-10, 1988, 
                                      Proceding edited by  K.Kleinknecht and E.Klempt,   volume 8, 1989, 

\bibitem{Haidenbauer_ZPA334_1989}    J. Haidenbauer, T. Hippchen, K. Holinde, and J. Speth, Z. Phys A 334, 467 (1989) 

\end{thebibliography}
\end{document}